\documentclass[12pt,preprint]{aastex}

\newcommand{\eg}  {e.g.,\;}        
\newcommand{\hub} {$H_{\mathrm{0}}\,$}

\newcommand{\kms}{\,km s$^{-1}$\,}

\newcommand{\velw}{$W_{20}$\ }
\newcommand{\be}[1]{\begin{equation}\label{#1}}
\newcommand{\ee}{\end{equation}}
\newcommand{\mpc}{\,Mpc\,$^{-1}$}

\shorttitle{The slope of the Baryonic Tully-Fisher relation}
\shortauthors{Gurovich et al.}

\begin{document}


\title{The slope of the Baryonic Tully-Fisher relation}
\shortauthors{Gurovich et al.}


\author{Sebasti\'an Gurovich \altaffilmark{1}, Kenneth Freeman, Helmut Jerjen, Iv\^anio Puerari\altaffilmark{2}}
\affil{Research School of Astronomy and Astrophysics, Mount Stromlo
  Observatory, Cotter Rd, Weston, ACT 2611, Australia}
\author{Lister Staveley-Smith \affil{School of Physics, University of Western Australia, Crawley, WA 6009, Australia}}


\altaffiltext{1}{Now CONICET fellow at the IATE, Cordoba, Argentina}
\altaffiltext{2}{Sabbatical visit from INAOE, Tonantzintla, Mexico}

\begin{abstract}
We present the results of a baryonic Tully-Fisher relation (BTFR)
study for a local sample of relatively isolated disk galaxies. We
derive a BTFR with a slope near $3$ measured over about $4$ dex in
baryon mass for our combined \textrm{H\,\scriptsize{I}} and bright spiral disk samples. This
BTFR is significantly flatter and has less scatter than the TFR
(stellar mass only) with its slope near $4$ reported for other samples
and studies.  A BTFR slope near 3 is in better agreement with the
expected slope from simple $\Lambda$CDM cosmological simulations that
include both stellar and gas baryons. The scatter in the TFR/BTFR
appears to depend on $W_{20}$: galaxies that rotate slower have more
scatter. The atomic gas--to--stars ratio shows a break near $W_{20} =
250$ \kms\, probably associated with a change in star
formation efficiency. In contrast the absence of such a break in the
BTFR suggests that this relation was probably set at the main epoch of baryon
dissipation rather than as a product of later galactic evolution.
\end{abstract} 

\section{Introduction}
\label{sec:intro}
The baryonic Tully-Fisher relation (BTFR) for disk galaxies relates
the total baryon disk mass to the disk rotational velocity
\cite[$\eg$][]{kcf99, mcg2000,bel01}.  It has long been recognized
that the (luminous) TFR implies a coupling between the luminous and
dark components of disk galaxies \citep[$\eg$][]{pie92}. Simple
cosmological arguments \citep[$\eg$][]{whi97} predict that the slope of the
BTFR should be close to 3.  In this approach, the galaxy mass is calculated within its virial radius, taken to be
the radius $r_{200}$ within which the mean baryon mass surface density
is $200$ times the critical density of the universe. In its simplest
form, the dark halo is modeled as a singular isothermal sphere with a
density distribution $\rho(r) = V^{2}/(4\pi G r^{2})$. The only
dimensional parameter is the rotational velocity $V$.  It follows that
$r_{200} = V/(10$\hub) where \hub is the Hubble constant, and the halo
mass within $r_{200}$ is $\mathcal{M}_{r_{200,\rm{halo}}} =
V^{3}/(10G$\hub).  If some fraction $f_{\rm d}$ of the halo mass is in
the form of gas which becomes the exponential disk of the galaxy, then
${M}_{\rm disk} = f_{\rm d} V^{3}/(10GH_{\mathrm{0}}\,)$. We would
then expect a BTFR with a slope of 3.  In this argument, the virial
radius within which the mass was estimated is not a structural
scalelength of the system in the sense of the scalelength of an
exponential disk: it depends on the rotational velocity.  This
predicted BTFR slope near $3$ is also seen in semi-analytic and
numerical simulations of galaxy formation within the $\Lambda$CDM
framework: see \citet{mo00, vand00, nav00, krav04}.

In reality, it appears that the rotational velocity $V$ of disk
galaxies depends on the gravitational fields of both the baryons and
the dark matter. $V$ is affected by the structure of the dark matter
halo, the initial angular momenta of the baryons and dark matter, the
structural evolution of the baryons and the adiabatic compression of
the halo by the disk. The stellar and gas baryon masses are affected
by baryon loss via winds and other feedback processes, the star
formation efficiency and history, all of which vary, possibly in a
systematic way, with galaxy mass and environment. Therefore, the
slope, zero-point and possible departures from linearity of the BTFR
should be sensitive to the many evolutionary processes that go on
during galaxy formation from the main epoch of hierarchical assembly
until the present time.  We should stress that all empirical TF/BTFR
studies suffer from the caveat that there is no way yet to measure
the rotational velocity at the virial radius, which may be larger
or smaller than the velocity inferred from $W_{20}$ measurements
\citep[e.g.][]{bat05}. Also, the baryonic mass may or may not be
proportional to the virial mass.

In this paper we derive the TFR and BTFR for a sample of relatively
isolated disk galaxies covering a large range in mass and rotational
velocity. Our rotational velocity measure is the width $W$ at twenty
percent of the peak of the integrated $\textrm{H\,\scriptsize{I}}$
profile. In \S 2 we describe the two samples of galaxies used in
this study.  In \S 3 we present the observations and data reduction,
and in \S 4 outline the method of our analysis. \S 5 contains the
results of the observed and derived quantities. In \S 6 our empirical
disk scaling relations are presented and some astrophysical
implications are discussed. In \S 7, we conclude with a summary of the
main results of this study. The Appendix includes an overview of the
results of similar studies by other authors along with some
discussion.

\section{The Sample Selection}
\label{sec:sam subsec:sam}

We use two local (D\,$<$\,130 Mpc) field galaxy samples for our BTFR
analysis: (i) the Sakai et~al. (2000) galaxies with Cepheid distances
for the bright end of the relations; (ii) our new
\textrm{H\,\scriptsize{I}} Parkes All Sky Survey (HIPASS) sample of
faint \textrm{H\,\scriptsize{I}}-selected galaxies.  This second
sample selected from the \cite{kil2002} and \cite{kor2004} catalogs
was re-observed and contains faint gas-rich galaxies with the
following selection rules: (1) none appears to be interacting; (2)
absolute magnitude brighter than M$_{\mathrm{\it{V}}} = -12.5$; (3)
$W_{20} < 290$ {km s$^{-1}$}, in order to overlap with the bright end
of the TF relation defined by the Sakai sample; (4) inclination $>
40^{\circ}$ (except one), to reduce errors in our derived velocity
widths from correcting for inclination; (5) galactic latitude $|b| >
20^{\circ}$ (except one); (6) optical diameter $< 2'$ to allow
single--pointing imaging in $H$--band with the CASPIR system on the
ANU 2.3--m telescope.

Because the HIPASS positions are imprecise, all optical IDs
of our HIPASS sources were verified by H$\alpha$ spectroscopy
with the ANU Dual Beam Spectrograph on the 2.3--m telescope.  The
spectroscopic observations showed that all the HIPASS sources were
correctly identified with the optical counterpart in the NASA
Extragalactic Database (NED) except HIPASS\,J1112-86 that is
misidentified in NED as the background galaxy ESO\,007-\,G\,004.

\section{Observations and Data Reduction}

The HIPASS catalog was our source for the \textrm{H\,\scriptsize{I}}
selected sample according to the selection rules of \S 2. However, the
HIPASS correlator (with $1024$ channels and $64$ MHz bandwidth)
provided a relatively poor velocity resolution of about $18$ \kms
\citep{bar01} which for a TF/BTF study of faint galaxies with line
widths \velw as low as 40 \kms was less than ideal. Therefore new
\textrm{H\,\scriptsize{I}} observations were obtained with higher
velocity resolution for our selected HIPASS galaxies, re--observed
with the Parkes 64--m telescope with pointed observations and the
narrow--band correlator ($1024$ channels and $8$ MHz bandwidth) that
offers a velocity resolution of about $6.6$ \kms\ after
processing. Given that the intrinsic \textrm{H\,\scriptsize{I}}
velocity dispersions within galaxies ($\sim$ 6\,--\,8\kms) are
comparable to the spectral resolution of the HIPASS $8$ MHz system,
the system was well suited to our needs.

The noise ($N$) in the \textrm{H\,\scriptsize{I}} spectra depend
on the resolution and integration time as \be{eqn:hisn} N \propto
\frac{1} {\left(t \Delta \nu \right )^{0.5}} \ee where $\Delta \nu$
is the observed frequency resolution and $t$ is the observing time,
so we increased the S/N at the higher resolution by combining several
individual observations with 24 cycles of 5 seconds of integration
per beam. This gave a total on-source integration time of 14 minutes
for the 7 inner HIPASS beams. We combined multiple observations for
all our galaxies (except IC\,5028) for which the total on-source
integration time typically exceeds 30 minutes, typically with a
S/$N \sim 10$ in peak \textrm{H\,\scriptsize{I}} flux.

The individual \textrm{H\,\scriptsize{I}} observations were processed
through the on-line reduction code LIVEDATA and multiple data sets
for a common source are gridded and combined using the program
GRIDZILLA, either at the telescope or post-processed. The MIRIAD
package was used for all subsequent \textrm{H\,\scriptsize{I}} data
analysis. The MIRIAD task \texttt{mbspect} was used to produce and
measure parameters of the \textrm{H\,\scriptsize{I}} spectra including
the integrated flux, the systemic velocity and the $W_{20}$ velocity
width. The \textrm{H\,\scriptsize{I}} spectra were separated from
several sources of interference both at the telescope and by
post-processing. For example, solar interference that occurred during
day-time observations generated baseline distortions which were
removed with the MIRIAD task \texttt{mbspect} which uses
an algorithm that minimizes the mean absolute deviation of the
flux-weighted velocities. Tests were performed to determine if the
measured velocity widths varied with the degree of hanning smoothing,
a parameter that is set in \texttt{mbspect}.

Our method for estimating the stellar mass from the light of galaxies
used $V$- and $H$-band photometric measurements, typically extending
over three disk scale lengths.  The $V$-band observations for our
\textrm{H\,\scriptsize{I}} selected sample were made with the ANU 1--m
and 2.3--m telescopes on several nights between 29.03.2001 and
24.06.2003 using two different detectors, the 8K $\times$ 8K WFI and
the single 2K $\times$ 2K imager CCD.  The WFI configuration used at
the f/8 Cassegrain focus of the 1--m telescope has 8 4K $\times$ 2K
3-side buttable CCDs arranged in a 2 $\times$ 4 mosaic, with 15 micron
pixels and a scale of 0.38 arcsec per pixel. The field of view is 1.2
degrees along the diagonal.  We also used the single 2K $\times$ 2K
thinned CCD at the same focus; it has 24 micron pixels, a scale of 0.6
arcsec per pixel and a field of view of 0.34 degrees along a side.
Some $V$-band images were taken at the Nasmyth focus of the 2.3--m
telescope, using the Imager focal reducer and a 1K $\times$ 1K CCD
with 24 micron pixels, a scale of 0.59 arcsec per pixel and a circular
field of view of 6.6 arcmin diameter.

The near-IR $H$-band observations used the ANU Cryogenic Array
Spectrometer/Imager (CASPIR) on the 2.3m telescope on several nights
between 05.02.2001 and 08.11.2003. The observations were made at
the Cassegrain f/18 focus, using a single 256 $\times$ 256 InSb
detector array with a field of view of 2 arcmin $\times$ 2 arcmin
and 30 micron pixels at 0.5 arcsec per pixel.  The total on-source
integration time for all our \textrm{H\,\scriptsize{I}} selected
galaxies typically exceeded 30 minutes for both the $V$ and $H$-bands
so that for some of our LSB dwarfs we reached as deep as 26.5 mag
arcsec$^{-2}$ in the $V$-band.

The standard packages of IRAF were used to reduce our $V$-band data.
The individual processed galaxy frames that were dithered during the
observations were registered by measuring the positions of common
stars in each frame. We removed any residual intensity offsets with
\texttt{imsurfit} before combining frames. A plane surface was fitted
to the sky: in most cases, a flat (constant) sky level gave the best
sky fit. We use the \cite{gra82} standards to derive the $V$-band
zero-points, typically accurate to 0.04 magnitudes.

The near-IR observations included sequences of science and sky
frames and the reduction was carried out with the \texttt{CASPIR}
package in IRAF. All the CASPIR frames were linearized except the
biases and darks.  A normalized flat frame was produced each night
and all linearized galaxy, sky, and IRIS standard star frames that
were observed on the same night were divided by this
normalized flat. Our $H$-band zeropoints are typically accurate to
0.05 magnitudes. The $V$-band galaxy frames were geometrically
transformed to match the pixel scale and orientation of the
corresponding $H$-band frame using the IRAF tasks \texttt{geomap}
and \texttt{geotran}.

Surface photometry measurements were made on the processed galaxy
images. We used the IRAF task \texttt{ellipse} to fit isophotal
ellipses to the pixel intensity distribution for each galaxy to
determine extrapolated magnitudes, scalelengths, ellipticities and
colors. The \texttt{ellipse} algorithm is described in \cite{jed87}.
Usually the $V$-band data has a higher signal--to--noise. Therefore in
all cases except one the isophotal fits were first made on the
combined $V$-band image and then the defined apertures (ellipses)
applied to the equivalent combined $H$-band image.  \texttt{Ellipse}
was run with the center of the first ellipse set to coincide with the
geometric center of the galaxy. The position angle and ellipticities
of the isophotes were then allowed to vary, keeping the centre fixed,
and the isophotes were examined with the IRAF task
\texttt{isoimap}. This is to confirm that the outer isophotes indeed
encompass the fainter parts of the galaxy in both the $H$ and
$V$-bands.  We adopted pure exponential disk models for all our
galaxies and excluded the inner few data points that in some cases may
include a small bar or bulge component. For the literature sample we
used published magnitude growth curves that model the galaxy light
distributions as pure exponential disk systems.  The integrated
magnitudes and surface brightness profiles were corrected for galactic
extinction using the extinction prescription of \cite{sch98}.

The $V$-band isophotal ellipticities provide a good estimate of
inclination \citep{mac2000}.  For our $\textrm{{H\,\scriptsize{I}}}$
sample we used the mean ellipticity of the three outermost fitted
isophotes from the $V$-band images, and adopted an intrinsic
minor--to--major axis ratio $q_\circ = 0.20$. Our inclinations were
derived using $\cos^2i = {[(b/a)^{2} -
    q_\circ\,^{2}]/(1-q_\circ\,^{2})}$ where $b/a$ is the isophotal
minor--to--major axis ratio with formal inclination errors of
typically 3 degrees. The axis ratio values for the Sakai galaxies are from
\cite{sak2000}.

Errors in the apparent magnitudes, integrated
\textrm{H\,\scriptsize{I}} fluxes, inclinations, $W_{20}$, distances,
       [Fe/H], and gas masses were propagated analytically from the
       errors in the observations, using conventional techniques. More
       details of the error analysis are described in \cite{gurofree}.

\section{Methodology}

We first calculate our TF and BTF relations using the stellar
$\mathcal{M}$/$L$ adopted directly from \cite{mcg2000} which we
present in Fig. \ref{tfbtf1mcg}. Then in Fig. \ref{tfbtf1} we
present our relations with the new stellar $\mathcal{M}$/$L$ obtained
from modelling the stellar population histories using the GALAXEV
stellar population synthesis (SPS) code of \cite{bc03}, described
below.

In summary, features of our TF/BTF study include: (1) distances
to galaxies without primary or secondary distances derived from
large-scale and Virgocentric flow models \citep[see,][]{gurofree}
scaled to \hub= 73 \kms \mpc; see \cite{gurofree} for details.  (2)
$\mathcal{M}$/$L$ ratio values derived from GALAXEV SPS models
\citep{bc03} using a bottom light IMF (Chabrier) and a star formation
rate that decays exponentially with a timescale constrained by at
least one color and metallicity derived from the metallicity--luminosity
relation of \cite{mat98}; (3) \textrm{H\,\scriptsize{I}} observations of the
\textrm{H\,\scriptsize{I}}-selected galaxies with the 64-m Parkes
narrow--band system; (4) detailed Monte Carlo propagation of errors
in observed quantities through the entire process to determine our
baryon mass errors as described in \cite{gurofree}; (5) use of
\cite{sch98} extinction values. 

\begin{figure}
    \plotone{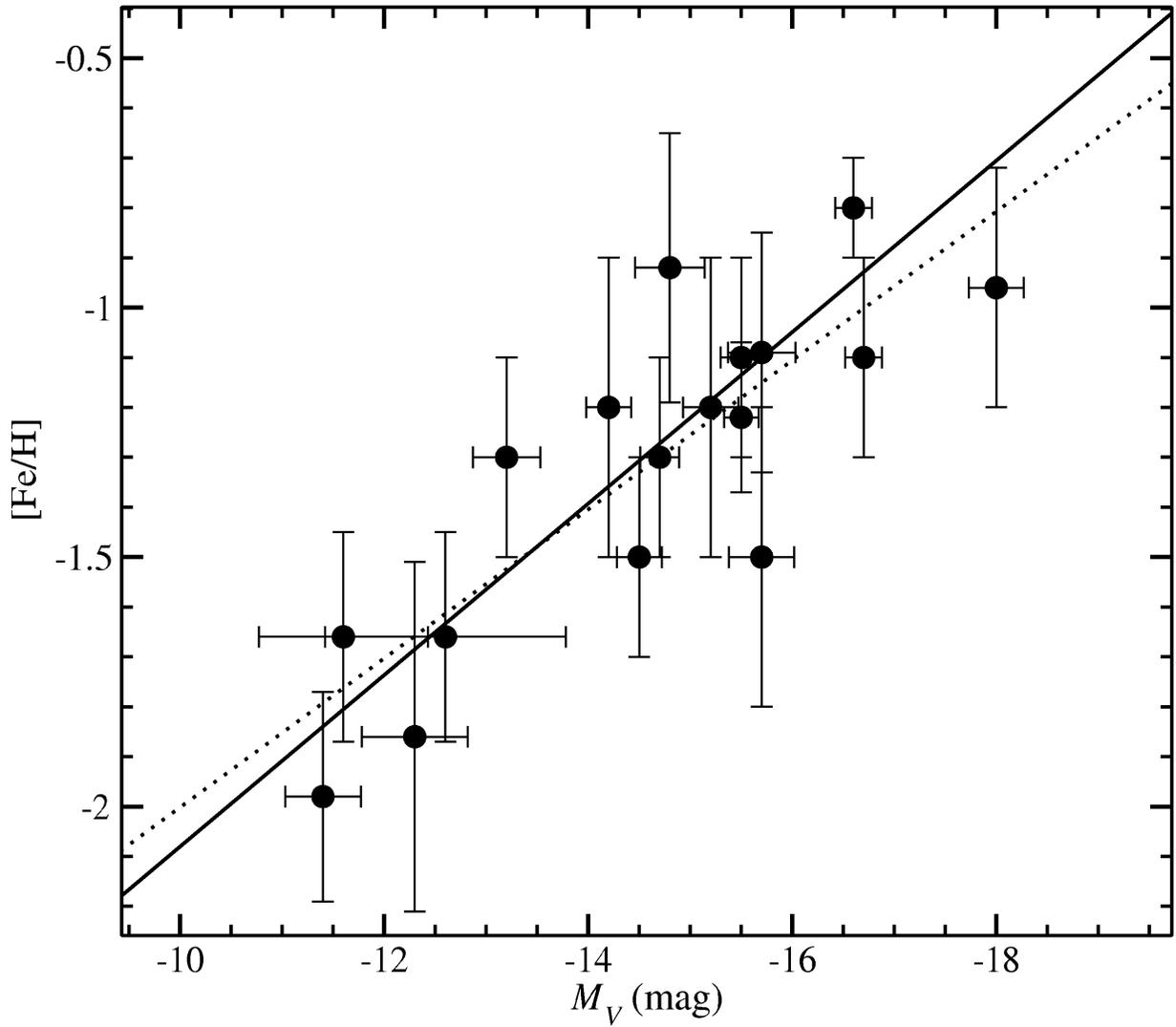}
  \caption{Adapted L-Z data for local group dwarf galaxies from Mateo
    98. The weighted bivariate fit (solid line) shown is used to interpolate this relation for our galaxies. \label{mateo_feh} }
\end{figure}

The stellar masses are determined for each galaxy with a Monte
Carlo implementation of the \cite{bc03} stellar population synthesis (SPS) code. The \cite{bc03} code
evolves the stellar population history of a single gas cloud of
uniform metallicity. Our goal is to estimate the $\mathcal{M}/L$
values (and errors) for all galaxies, constraining them with the
observed colors and adopted metallicities.  The color and metallicity
values also have known errors and so for each galaxy we run a set of
SPS models, choosing the color and metallicity from the gaussian
adopted probability distributions associated with the errors. This
generates a probability distribution of $\mathcal{M}/L$ values for
each galaxy.  We choose the age ($i.e.,\;$ the lapsed time since star
formation began) that reproduces the observed galaxy
colors, restricting them to be between 8 and 13.75 Gyrs to be
consistent with the observed ages of the old disk stars in Local Group
galaxies \citep[e.g.,][]{mat98}.  The metallicity for each galaxy is
adopted from the literature (when available), or determined by
interpolating a weighted bivariate linear fit: $\rm{[Fe/H]} = a + b
M_V$, to the \cite{mat98} L-Z dwarf data with FITEXY \citep{pres92}.  Fig. \ref{mateo_feh} shows
the Mateo L-Z data with weighted fit that we use to estimate the error
in our interpolated [Fe/H] values using Equation \ref{eqn:mateo}.  The
$\sigma_{ab}$ covariance term of Equation \ref{eqn:mateo} is negative
and similar in magnitude to the sum of the three positive terms so the
expected error in our metallicity values is small.  For each galaxy,
the [Fe/H] error values are set to be the standard deviation of the
[Fe/H] probability distribution for the Monte Carlo stellar population
synthesis simulations. The errors in our [Fe/H] values are calculated
from the errors in the M$_V$ values and from the calculated parameters
of the fitted L-Z relation, where, $a=-3.8$, $b =-1.7\, \times 10^{-1}$,
$\sigma_{a}^{2} = 1.4\,\times 10^{-2}$, $\sigma_{b}^{2} = 6.0 \,\times
10^{-4}$ and $\sigma_{ab} = -9.2\,\times 10^{-3}$, and:

\be{eqn:mateo}\hspace{-0.001in}\sigma [\frac{Fe}{H}]^{2} =
\sigma_{a}^{2} + M_{\textrm{v}}^{2}\sigma_{b}^{2} + b^{2}
\sigma_{M_{\textrm{v}}}^{2} + 2M_{\textrm{v}}\sigma_{ab} \label{eqn:mateo}
\ee\\

The SFR of our models is: $\psi(t)= 1[\mathcal{M}_{\odot} + \epsilon
  \mathcal{M}_{\textrm{PG}}(t)]\tau^{-1}\exp(-t/\tau)$, where
$\mathcal{M}_{\textrm{PG}}(t)= 1\mathcal{M}_{\odot}[1-\exp(-t/\tau)] -
\mathcal{M}_{\textrm{stars}} - \mathcal{M}_{\textrm{remnants}}$ is the
mass of gas processed into stars and returned to the ISM and
$\epsilon$ is the fraction of this gas allowed to recycle into new
star formation. In our models, we do not include gas recycling or 
infall, so we set $\epsilon = 0$.  We make this simplification
because we wish to treat our galaxies uniformly, even though this
assumption is more likely to be true for dwarfs with shallower
potential wells than for spirals. 

Each galaxy is modelled to cover a wide range of possible star
formation history with a Chabrier IMF and star formation rate (SFR)
that decreases exponentially in time. For each galaxy we consider star
formation histories with different e-folding timescales, selecting
values between $\tau$=0.1 and 14 Gyr. Larger values of $\tau$ have a
nearly constant SFR history and smaller $\tau$ values are close to a
single burst history. The observational constraints to the BC03 code
are the input colors and [Fe/H] values. The errors on colors and
[Fe/H] values are determined analytically from the observed and
derived parameters, and are assumed to have gaussian probability
distributions \citep{gurofree}. Linear interpolations were made over
the BC03 output grid of discrete [Fe/H], color, $\mathcal{M}/L$, and
age values, because the `observed' input values usually do not fall on
the BC03 grid. To calculate the errors in the output $\mathcal{M}/L$
values at a given $\tau$, the simulations for each galaxy were rerun
about 100 times, each time using a new input pair of values for color
and [Fe/H] selected from their gaussian error distributions.  In this
way, a distribution of $\mathcal{M}/L$ value was produced for each
galaxy.  The mean of the $\mathcal{M}/L$ values over all runs, given
the age restriction described above, is chosen and the $\mathcal{M}/L$
error is taken to be the standard deviation about the mean of all
$\mathcal{M}/L$ values at a given $\tau$. For each galaxy this process
is repeated for the range of $\tau-$values, as above, and the mean of
all $\mathcal{M}/L$ values is adopted to be our final $\mathcal{M}/L$
value with error given by the standard deviation about the mean of all
$\mathcal{M}/L$ values.

The SPS models for our \textrm{H\,\scriptsize{I}} selected galaxies
are constrained with ($V-H$) colors measured at the half-light radius
because the half-light photometry is more precise than at larger
radii. The apparent magnitudes and colors are corrected for galactic
extinction following \cite{sch98} and shown in Tab.
\ref{tab:hiphot_obs}.  The $V$ and $H$-band photometry give two partly
independent estimates of the stellar mass. The total luminosity of the
galaxy in each band is derived by extrapolating its radial surface
brightness profile, and the stellar mass is then estimated using the
derived $\mathcal{M}$/$L$ values from the SPS simulations. Even though
the theoretical $H$-band $\mathcal{M}/L$ values are probably better
estimates of the true $\mathcal{M}/L$ values because of the added
sensitivity towards the older stellar populations and the fact that
near-IR bands are less effected by extinction, in practice the
$H$-band surface photometry is not as deep, due to the inefficiency of
near-IR detectors. Therefore, as a compromise between induced
photometric errors and SPS model errors, we choose the weighted mean
of the stellar mass values, calculated using $\mathcal{M}$/$L$ and
luminosity values for a minimum of two bands, so the adopted disk mass
(stellar) is the weighted mean of the $V$ and $H$-band disk masses.

Similarly, the stellar mass estimates for the Sakai galaxies are
determined by constraining the stellar population histories using
($V-H$) and ($V-I$) colors and with metallicities obtained from the
literature, with references shown in Tab. \ref{tab:dis_sak}. The
adopted stellar masses for these galaxies are taken to be the
weighted mean of all four stellar mass estimates except for NGC\,3319
and NGC\,4548 for which only two $\mathcal{M}$/$L$ values could be
calculated.

The gas (atomic) masses are determined following equations
(\ref{eqn:HIa}) and (\ref{eqn:HIb}) to account for He and the heavier
elements. \be{eqn:HIa} \mathcal{M}_\textrm{{H\,\scriptsize{I}}}\; =
2.36 \times 10^{5}D^{2} F_{\textrm{H\,\scriptsize{I}}\;} \ee where
$F_{\textrm{H\,\scriptsize{I}}}$ is in Jy\,\kms and $D$ is in Mpc.
\be{eqn:HIb} \mathcal{M}_{\textrm{{gas}}} = 1.4 \times
\mathcal{M}_\textrm{H\,\scriptsize{I}}. \ee Inclusion of the available
H$_{2}$ gas component for 6 of the Sakai galaxies did not have a
significant effect on the TF/BTFRs; because molecular gas masses are
not available for the HIPASS sample, we do not include molecular gas
in our baryonic masses.

\section{Results}
\label{sec:results}

The integrated \textrm{H\,\scriptsize{I}} flux values for the Sakai
galaxies are tabulated in Tab. \ref{tab:sakhi}. These fluxes are not
listed in \cite{sak2000} but come from several sources in the
literature. The weighted mean of the literature values is calculated
and found to have a typical error of 25\%. The distances to the Sakai
galaxies are from \cite{sak2000} and the metallicities for the Sakai
galaxies are taken from the literature. Along with the derived gas
mass and luminosity values, the reference list is presented in Tab.
\ref{tab:dis_sak}. The \velw values for the Sakai galaxies
(corrected for inclination and redshift) are from table 2 of
\cite{sak2000}. A redshift correction is not included for our other
two galaxy samples because the effect on \velw is negligible within
our redshift range.  Tables \ref{tab:dis_sak},
\ref{tab:dis_mlt_sak_vmi}, \ref{tab:dis_mlt_sak_vmh} and
\ref{tab:sak_mas_den} list the derived quantities for the Sakai
galaxies including the [Fe/H], stellar $\mathcal{M}$/$L$s, stellar
masses, gas--to--star fractions and mean baryon mass surface density values
(see \S \ref{subsec:sigma}). Tables \ref{tab:hiphot1} and \ref{tab:hiphot_obs} list
the surface photometry measurements for our \textrm{H\,\scriptsize{I}}
selected galaxies. Other measured and derived optical and
\textrm{H\,\scriptsize{I}} quantities can be found in Tables
\ref{tab:pksnarrow} and \ref{tab:dis_hipass}.  Our
\textrm{H\,\scriptsize{I}} flux values are in good agreement with the
HIPASS literature values, our typical v$_{\mathrm{helio}}$ errors are
about $5$\kms\ and our \velw measurements have typical errors between 5 and
10\kms. The baryon mass values including gas--to--star ratio and mean baryon mass
surface densities are also tabulated in Tables \ref{tab:dis_mlt} and
\ref{tab:mas_den} and used for our relations in subsequent figures.

We compare our $\mathcal{M}$/$L$ for our two galaxy samples with the
range of modelled $\mathcal{M}$/$L$ with Scaled Salpeter IMF found in
table 3 of \cite{bel01}, for which we constrained 297 and 531 models
with our ($V-H$) and ($V-I$) colors. As is evidenced in
Fig. \ref{f_sps} our $\mathcal{M}$/$L$ values coincide with the higher
probability density bins of the \cite{bel01} output $\mathcal{M}$/$L$
grid (in contours). Our models therefore, which include color, IMF,
ages, and [Fe/H] constraints, produce $\mathcal{M}$/$L$ values that
are consistent with those adopted by \cite{bel01} in their scaled
Salpeter IMF models. The same IMF was used to construct their
preferred BTFR. We note that systematically higher $\mathcal{M}$/$L$
values that could result from a different universal IMF scaling will
act to steepen a BTFR, since the effect of any differential
gas--to--star fraction trend with \velw, would be diluted. We explore
this effect by scaling our modelled $\mathcal{M}$/$L$ later in Sec.
\ref{sub:slope}.

The errors in the baryon mass values depend on the errors in (i)
$\mathcal{M}$/$L$ (obtained from the stellar population synthesis
models), (ii) apparent magnitudes, (iii)
\textrm{H\,\scriptsize{I}} fluxes, and (iv) flow model distances. The
$\mathcal{M}$/$L$ errors themselves include errors in the observed
colors and metallicities; errors from the limitations of the stellar
population history models have not been included.  For galaxies
without primary distances, the distance errors are the largest
fractional error contribution and can provide up to 30\% of the
total baryon mass error. Next are the errors in the $\mathcal{M}$/$L$,
then the \textrm{H\,\scriptsize{I}} flux errors, and finally the
apparent magnitude errors provide the smallest contribution to the
baryon mass error budget. See \cite{gurofree} for more details. We
note that the analysis for the Sakai galaxies shows that the stellar
masses are only weakly sensitive to the the color constraint chosen,
so the chosen color does not significantly alter the slope of the
BTFR. This is consistent with what is found by \cite{bel01}.

\section{TF/BTFR}
Our TF/BTFR spans four orders of magnitude in stellar mass, from $1.1
\times 10^{7}$ to $1.6 \times 10^{11}$ $\mathcal{M}_\odot$. Both the
baryon mass and the $W_{20}$ values have errors, so we determine
weighted bivariate fits with FITEXY from \cite{pres92}, as well as
weighted forward and reverse linear least squares fits. We present two
versions of the TF/BTFR. The first is derived using the constant
stellar $\mathcal{M}$/$L$ values directly from \cite{mcg2000} and the
second uses our SPS stellar $\mathcal{M}$/$L$ values: see Figs.
\ref{tfbtf1mcg} \& \ref{tfbtf1} respectively. A comparison can be
found in Tab. \ref{tab:tf_fits}. In both cases, separate fits are
shown for each sample, and for the combined Sakai and
\textrm{H\,\scriptsize{I}} samples. Qualitatively the fits in both
figures are similar.  When the \textrm{H\,\scriptsize{I}} gas is
included, the BTFRs become flatter and the scatter is reduced. Our
fits do not change significantly when we correct our $W_{20}$ values
for an \textrm{H\,\scriptsize{I}} velocity dispersion of $6$ km/s,
following equation 1 of \cite{swa2003}, so we do not correct for the
velocity dispersion of the gas. We note for comparison that forward
and reverse BTFR fits for the Sakai galaxies computed by a least
squares routine that derives the error in slope from the weighted
residuals of data points about the fitted line in $\mathcal{M}$ and
separately in \velw are $2.9\pm 0.3$ (forward) and $3.1 \pm 0.3$
(reverse) and for the combined sample both the forward and reverse
BTFR weighted fits are calculated to have slopes of $3.1 \pm 0.1$. If
we remove NGC\,1365, the most massive outlier galaxy, then the
weighted BTF slopes for the Sakai sample alone are $2.7 \pm 0.3$
(bivariate), $2.6 \pm 0.1$ (forward) and $2.7 \pm 0.1$ (reverse).  The
BTFR slopes for the combined sample, with or without NGC\,1365, are
all (bivariate, forward and reverse) calculated to be $3.1 \pm 0.1$.
The combined sample appears to provide a consistent estimate of the
BTFR slope.

We find that the scatter in the TFR/BTFR is larger for galaxies at
lower $W_{20}$ and compare the scatter in Fig. \ref{tfbtf1} for
galaxies with $1.8 < $ log $W_{20}$ (km/s) $<2.2$ and $2.4<$ log
$W_{20}$ (km/s) $<2.8$ by calculating the ratio of the reduced
$\chi^2$ values (defined in the usual way) for these two
intervals. For the TFR and BTFR the ratios are $2.6$ and $1.6$
respectively.

Our \textrm{H\,\scriptsize{I}} selected sample is weakly biased
towards systems of lower $W_{20}$, which are detected at a higher S/N
for a given \textrm{H\,\scriptsize{I}} flux. The \cite{sak2000}
galaxies are mostly large, nearby, modeled with multiple colors and so
are relatively free of systematic effects.

The accurate Cepheid distances for the Sakai sample lead to a tight
stellar mass TFR with a slope of $4.3 \pm 0.4$, using our SPS models.
This is similar to the typical $H$-band luminosity TFR slope ($4.4 \pm
0.3$) and confirms that the $H$-band luminosity is indeed a fair
measure of the stellar mass in massive disks.  The BTFR for the Sakai
galaxies is flatter and even tighter than the TFR, with a slope of
$3.1 \pm 0.3$ (see Tab. \ref{tab:tf_fits}), close to the value
expected from cosmological arguments.  This result, that even for the
brighter galaxies, the \textrm{H\,\scriptsize{I}} mass fraction
increases sufficiently with decreasing $W_{20}$ to cause the marked
flattening between the TFR and the BTFR was somewhat surprising and is
discussed later in the paper. The BTFR slopes for the
\textrm{H\,\scriptsize{I}} selected sample, the Sakai sample, and the
two samples together are all similar and close to $3$.  

\subsection{Gas Fraction}
With the stellar mass estimates from our population synthesis analysis
and our \textrm{H\,\scriptsize{I}} mass values, we show in Fig.
\ref{gasfrac1} how the gas--to-stars ratio ($r_{\rm g}$) defined as
$\mathcal{M}_{\textrm{gas}}/\mathcal{M}_{\textrm{stars}}$, varies with
$W_{20}$.  The errors in $r_{\rm g}$ for the two samples are similar
and relatively small because the ratios are distance-independent.  The
large scatter for galaxies with lower $W_{20}$ appears to be real and
is presumably a consequence of variable star formation efficiency. The
brighter disk galaxies show rapidly decreasing $r_{\rm g}$ values with
increasing $W_{20}$ (see also \cite{geha2006}.  The transition from
the decreasing $r_{\rm g}$ values for the bright galaxies to the
roughly constant values for the fainter galaxies occurs near
$\log{W_{20}}$ (km/s) = $2.4$. This transition value, which appears to
mark the change in star formation efficiency, is discussed below.

\subsection{Baryon Surface Density}
\label{subsec:sigma}

At Joe Silk's suggestion, we used the stellar masses from our SPS
models to examine how the mean baryon mass surface density
$\Sigma_b$, depends on $W_{20}$.  We have only integrated
\textrm{H\,\scriptsize{I}} masses for our galaxies, and so assume that
the \textrm{H\,\scriptsize{I}} in each galaxy has an exponential
distribution with scalelength equal to the optical scalelength ($h$).
The adopted optical $h$ is taken to be the mean of the $V$ and $H$-band, $h$ for
the \textrm{H\,\scriptsize{I}} selected sample and our
re-calculated $V$-band $h$ values from the brightness profiles of
\cite{mac2000} for the \cite{sak2000} galaxies.  The mean baryon mass surface
density estimator is then $\Sigma_b = \mathcal{M}_{\rm baryons} /2 \pi
h^{2}$. Fig. \ref{denvsw20_1} shows that the surface density $\Sigma_b$ has a roughly linear dependence on \velw.

In estimating $\Sigma_b$, we assume that the
\textrm{H\,\scriptsize{I}} and light have the same scalelength.
However, it is likely that the ratio of the \textrm{H\,\scriptsize{I}}
size of the galaxy to its optical size increases towards fainter
systems. This would tend to somewhat steepen the slope of the
$\Sigma_b - W_{20}$ relation. To explore this further, we modelled the
effect for the sum of two exponential disks: stellar and
\textrm{H\,\scriptsize{I}}.  We assume from Fig. \ref{gasfrac1} that
the stellar and \textrm{H\,\scriptsize{I}} mass values for a given
dwarf are equal, take the ratio of scalelengths
$h_{\textrm{H\,\scriptsize{I}}}$/$h_{\star} =\gamma$, and the mean
surface brightness inside the half-light radius to be
$\rm{R}_{\rm{e}}$, a parameter commonly used for more general surface
brightness distributions. Tab. \ref{tab:logsigma} shows how
$\rm{R}_{\rm{e}}$/$h_{\star}$ and $\Sigma_{b}$ within
$\rm{R}_{\rm{e}}$ change with $\gamma$. This analysis shows that if
$\gamma$ increases as $W_{20}$ decreases, then we would get an
increase in the slope of $\log \Sigma_b$ \textit{vs} $\log W_{20}$
from $\sim 1$ to $\sim 1.3$.

\section{Summary and Conclusions}
\label{sec:conc}
We choose a sample of isolated disk galaxies ranging from faint dwarfs
to bright spirals. We construct TF and BTF relations and explore the
difference between the theoretically predicted BTFR slope of 3 and the
TFR slope of 4 obtained by many observers.  Regarding this difference,
\cite{vand00} argued that ``the physics regulating star formation and
feedback, coupled with the mass dependence of halo densities and
stellar populations has to tilt the TF relation to its observed slope.
The introduction of a stability-related star formation threshold
density increases the slope of the TF relation $\ldots$''. Our results
are entirely consistent with this argument. As \velw decreases, the
increasing gas--to--stars ratio and decreasing mean baryon mass
surface density, possibly associated with a decreasing trend in star
formation efficiency, generate the ``tilt" between the TFR and the
BTFR.

We show the gas--to--stars ratio for our combined sample, and the
break at baryon masses near $1$ x $10^{10}
\mathcal{M}_{\odot}$. Because the BTFR shows no such break, and its
slope is close to that expected from cosmological arguments, one could
argue that the total baryon content of isolated disk galaxies (as
measured by stellar + \textrm{H\,\scriptsize{I}} mass) has not
been much affected by galaxy evolution, including star formation history. In
this sense, the BTFR would be a fundamental relation relating back to
the main epoch of galaxy assembly.

There are some systematic uncertainties which affect any discussion of
the BTFR. (1) A problem inherent to any TF study is the change in
\textrm{H\,\scriptsize{I}} profile shape with $W_{20}$ \citep{nor07}
and therefore an uncertainty in how to relate $W_{20}$ to the
rotational velocity $V$ across the whole range of $W_{20}$ values.
(2) We have not included ionized or molecular gas in the total
baryonic masses. The idea of large amounts of molecular gas in the
dwarfs seems unlikely \cite[$\eg$][]{pil04,rea2005}. However, a larger
fraction of ionized undetected baryons in the more massive galaxies
would steepen the slope of the true BTFR. This ionized (warm) gas in
the more massive galaxies \cite[$\eg$][]{mall04, fuku06} may turn out
to be more significant in this respect.

\section{Appendix: Recasting the TFR}

\cite{aar83} were amongst the first to observe a bandpass-dependent luminous
TFR slope, which has been determined to steepen from
$\sim 3$ in the blue to $\sim 4$ towards the near-IR
\citep[$\eg$][]{sak2000}. The search was soon on for a band-pass independent
slope that attempted to include all baryons. In the Appendix, we
discuss the range of published BTF (gas + stars) slopes available in
the literature and we attempt to explain the reason behind some of
the discrepancies in these measured values.

However, the fitted BTF and TF parameters that result from most
observational studies do not include confidence intervals, determined
from robust statistical methods, so the task of comparing parameters
quantitatively is not attempted here.  Instead, our aim is to
qualitatively compare the various BTF slopes thus far obtained.

Several approaches have been pursued to recast the luminous (stars
only) TF relation to one which also includes the gaseous disk mass
component. The most common is to use baryonic mass instead of stellar
luminosity \cite[$\eg$][]{mcg2000,bel01}. In this approach, the mass
of the stars is estimated by using a fixed stellar $\mathcal{M}/L$
\cite[$\eg$][]{mcg2000,nor07}, or by modelling the luminous component
with stellar population synthesis codes, $\eg$ \cite{bel01}, or by
using mass models to fit rotation curves,
\cite[$\eg$][]{McG2005}. Another approach is to notionally convert the
atomic gas to a luminosity ($\eg$ \cite{kcf99,ver2001}). Each method
has its strength and weakness. For example, the assumption of a
constant stellar $\mathcal{M}/L$ is simple, but it is known that the
history of star formation is not uniform from one galaxy to the next
and that stellar mass and luminosity of galaxies are sensitive to
stellar population history effects \citep{bel01}. Converting gas to
stars again requires adoption of an appropriate $\mathcal{M}/L$ ratio.
Mass modelling, on the other hand is only possible for those galaxies
with well measured (and behaved), \textrm{H\,\scriptsize{I}} rotation
curves that extend out to large radii.

\subsection{The slippery BTFR slope}
\label{sub:slope}

In this section, we briefly discuss the measured BTFR slopes
determined by various recent authors. Although some studies report
slopes closer to 4, the reported slopes are mostly in the range $3.1$
to $3.7$ and most authors find that the BTFR is significantly flatter
than the the TFR (stars only). At least some of the differences in
slope come from different methodologies, including use of different
kinematic indicators as well as assumptions about the stellar
population histories used to calculate the $\mathcal{M}/L$ ratios. So here,
we report on some of the results thus far:

\cite{bel01} use stellar population synthesis models to derive a
slope (unweighted) for brighter disk galaxies of $3.51 \pm 0.19$,
and argue that the BTFR slope would be even flatter if fainter disks
were sub-maximal (as is widely believed). 

In a study of disk galaxies in the Ursa Major cluster, \cite{ver2001}
converts gas to luminosity and derives a range of BTFR slopes, between
3 and 4 for different gas conversion values $\mathcal{M}_{\rm
  gas}$/$L_{\rm K'}$. He finds that a BTFR slope of 4 gives the least
scatter when a constant $\mathcal{M}_{\rm gas}$/$L_{\rm K'}= 1.6$ is
adopted. This result is based on calculating the reduced $\chi^2$
values for different values of $\mathcal{M}_{\rm gas}$/$L_{\rm
  K'}$. We note however that the reduced $\chi^2$ values in this
analysis do not appear to pass through a minimum, so it is difficult
to judge the significance of this result. Moreover, recent results by
\cite{nor07}, using the more common methodology of converting
luminosity (K'-band) to baryon mass with a fixed stellar
$\mathcal{M}$/$L$, for mostly the same galaxies yield a markedly
flatter BTFR slope of $3.36 \pm 0.1$ and $3.04 \pm 0.08$ when the
asymptotic rotational velocity and $W_{20}$ are respectively used as
the kinematic rotation value. This appears to be statistically
significant and but in itself is a selection effect which we discuss
towards the end of the paper.

\cite{kas2007} on the other hand use a different method that uses a
kinematic estimator which accounts for disordered or non-circular gas
motions. They obtain a `TFR' slope of $3.3 \pm 0.2$ (lower panel of
their fig 1) at low z and a fitted `BTFR' slope slightly flatter than
3. It is noted however that some of their low mass galaxies are
classified as disturbed or compact systems and that there may be a
systematic effect since rotation curves from emission line data are
likely to be still growing at the last measured point.

\cite{geha2006} find an extremely flat BTFR slope of $1.89 \pm 0.08$
for their sample of SDSS dwarf galaxies. However when combined with
brighter samples from the literature that include the \cite{ver2001}
sample, the they find a BTFR slope of $3.70 \pm 0.15$, consistent with
the BTFR LCDM slope predicted by \cite{bul01} of $3.4\pm 0.1$, similar
to that found by \cite{rij2007} who obtain a BTFR slope of $\sim 3.2$
using stellar population synthesis modelling to estimate the stellar
mass component of their disks. When stellar population synthesis
models are used to calculate the stellar mass component of galaxies,
the derived BTFR slopes determined by authors are generally flatter
and typically closer to the value of $\sim 3$ expected from
$\Lambda$CDM cosmological simulations. However, some authors, do find
slopes closer to $4$. \cite{starketal2009} for example use rotation
curve data to create a BTFR and argue that line widths are not
accurate enough. Their sample consists of gas rich galaxies of
intermediate baryon mass for which the flat part of the rotation curve
is reached. They calculate BTFR slopes that are not much effected by
the chosen stellar population model (IMF). \cite{tra09} obtain a BTFR
slope consistent with 4 for a small sample of low mass galaxies but
given the small baseline in rotational velocity, and that many of
their rotation curves appear to still be rising at the last measured
point, their data may also be consistent with a significantly
different BTFR slope. 

\cite{meyeretal2008} use $\textrm{H\,\scriptsize{I}}$ velocity width
data and stellar $\mathcal{M}$/$L$ from stellar population synthesis
modelling that include $\mathcal{M}$/$L$ values from Bell \& de Jong
(2001) as well as a recipe for gas mass that includes a molecular
contribution. They find a slope near 4 but only for galaxies with
maximum rotational velocities $> 100$ km/s: see their fig. 14. These
relatively massive galaxies would mostly have
$\textrm{H\,\scriptsize{I}}$ extending out to the flat part of the
rotation curve. On the other hand, \cite{McG2005} also find a flat
BTFR with slope of $3.37 \pm 0.13$ with stellar population synthesis
models, without an additional IMF scaling. However \cite{McG2005} also
compute significantly flatter BTFR slopes for their maximum stellar
IMF scaling as well as with their maximum disk and their MOND modelled
baryon masses and also report BTFR slopes $\sim 4$ (see their tab. 2).

\begin{figure}
    \plotone{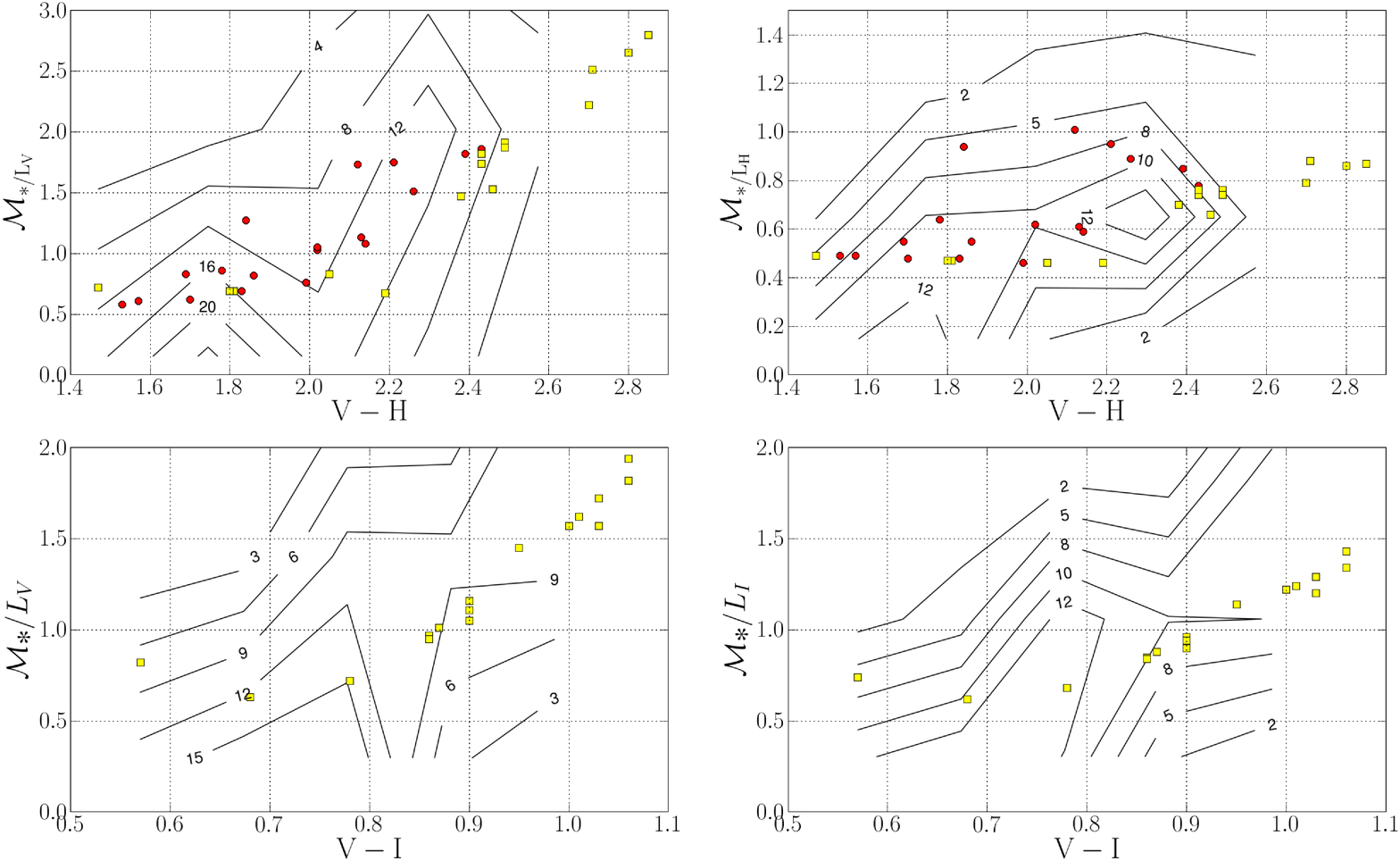}
  \caption{$\mathcal{M}$/$L$ \textit{vs} color for the Sakai (yellow
    squares) and \textrm{H\,\scriptsize{I}} (red circles) samples with probability density of the Scaled
    Salpeter models (contours), from table 3 of \cite{bel01},
    constrained with our ($V-H$) and ($V-I$) colors \label{f_sps} }
\end{figure}

\begin{figure}
    \plotone{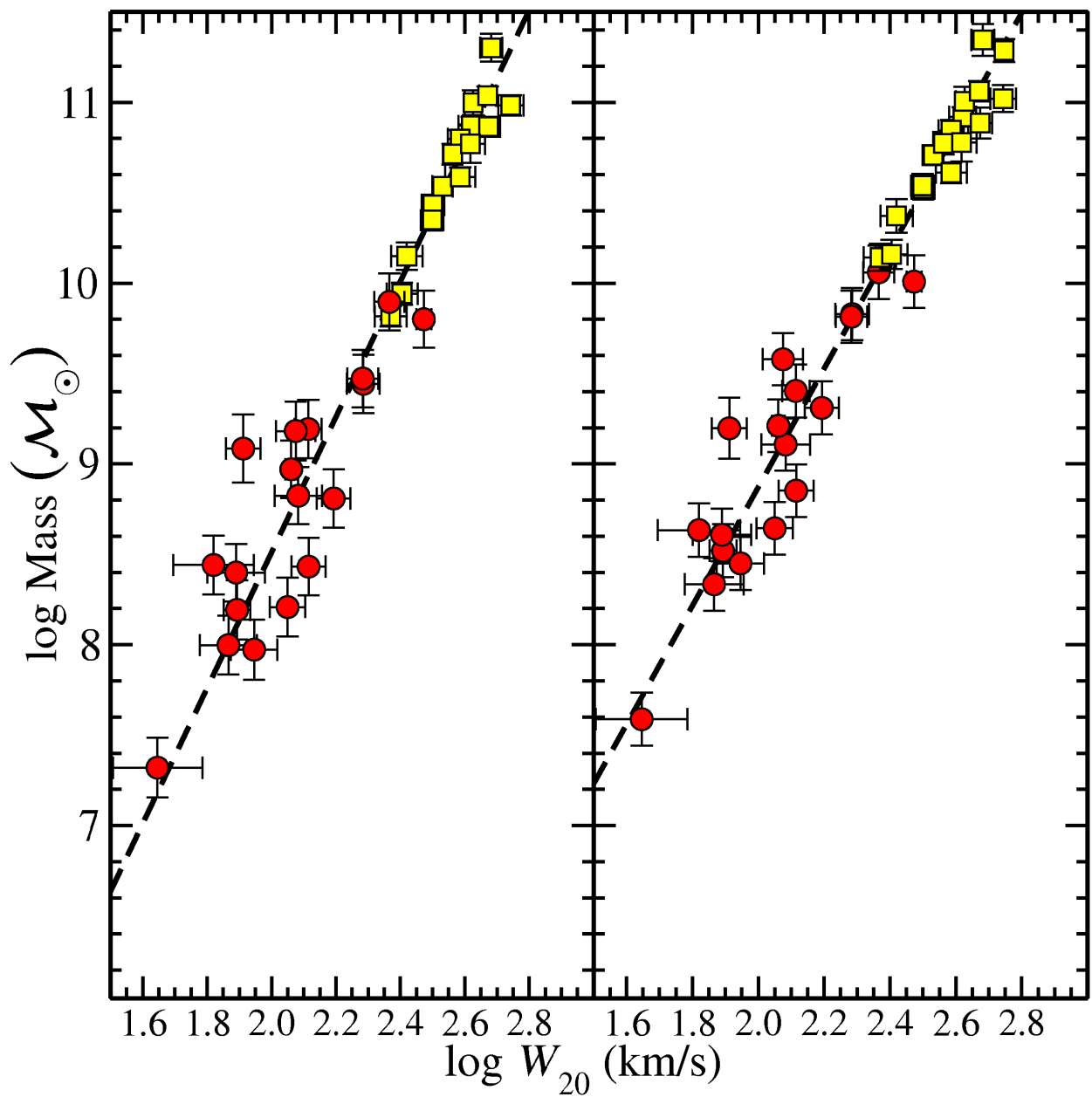}
  \caption{TF (left panel) and BTF (right panel) relations for the two samples of disk galaxies. Stellar masses (left panel) are shown with $\mathcal{M}$/$L$ values  taken from \cite{mcg2000} for the \textrm{H\,\scriptsize{I}} selected (red circles), and
  \cite{sak2000} (yellow squares) with weighted bivariate fits
  (solid line). Weighted bivariate fits for the union of the \textrm{H\,\scriptsize{I}} and \cite{sak2000} galaxies are also shown (broken line).  \label{tfbtf1mcg} }
\end{figure}

\begin{figure}
    \plotone{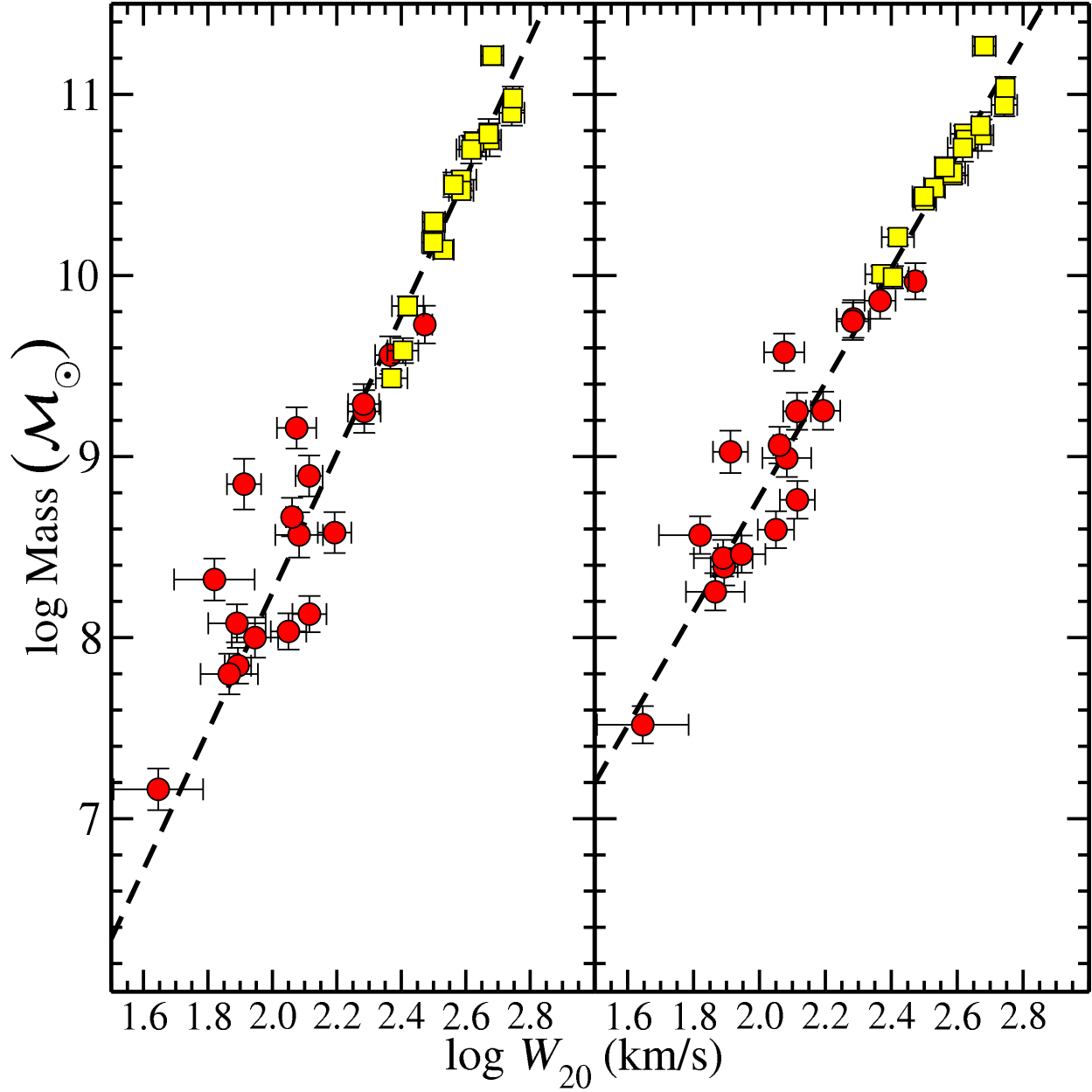}
  \caption{Same as Figure \ref{tfbtf1} but with stellar
    $\mathcal{M}$/$L$ values calculated from the simple stellar
    population synthesis models of \cite{bc03} with weighted bivariate
    fits (broken line). \label{tfbtf1}}
\end{figure}

\begin{figure}
  \plotone{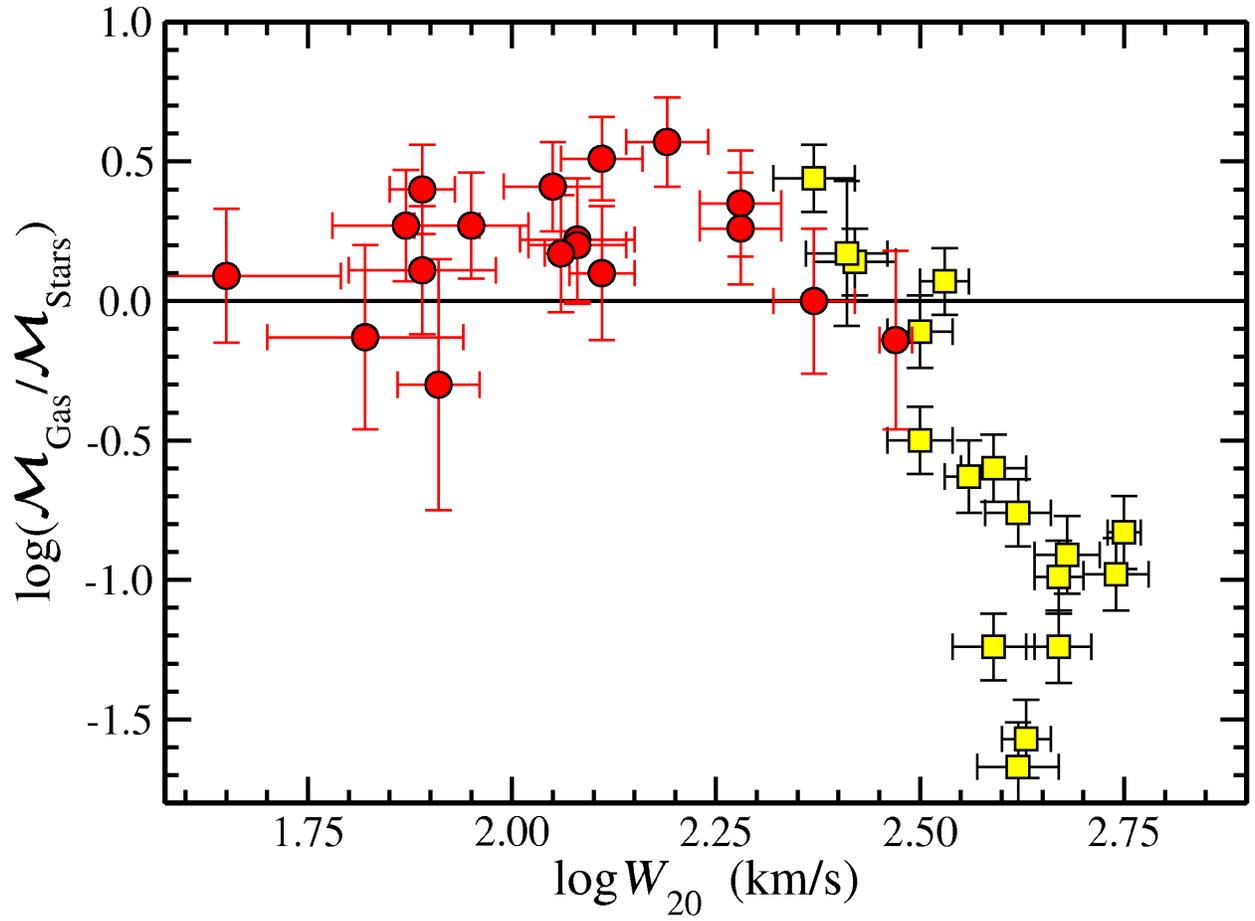}
  \caption{$\mathcal{M}_{gas}$/$\mathcal{M}_{stars}$
  fraction \textit{vs} $W_{20}$ for the two samples of disk galaxies: \textrm{H\,\scriptsize{I}} selected (red circles), and
  \cite{sak2000} (yellow squares). \label{gasfrac1}}
\end{figure}

\begin{figure}
  \plotone{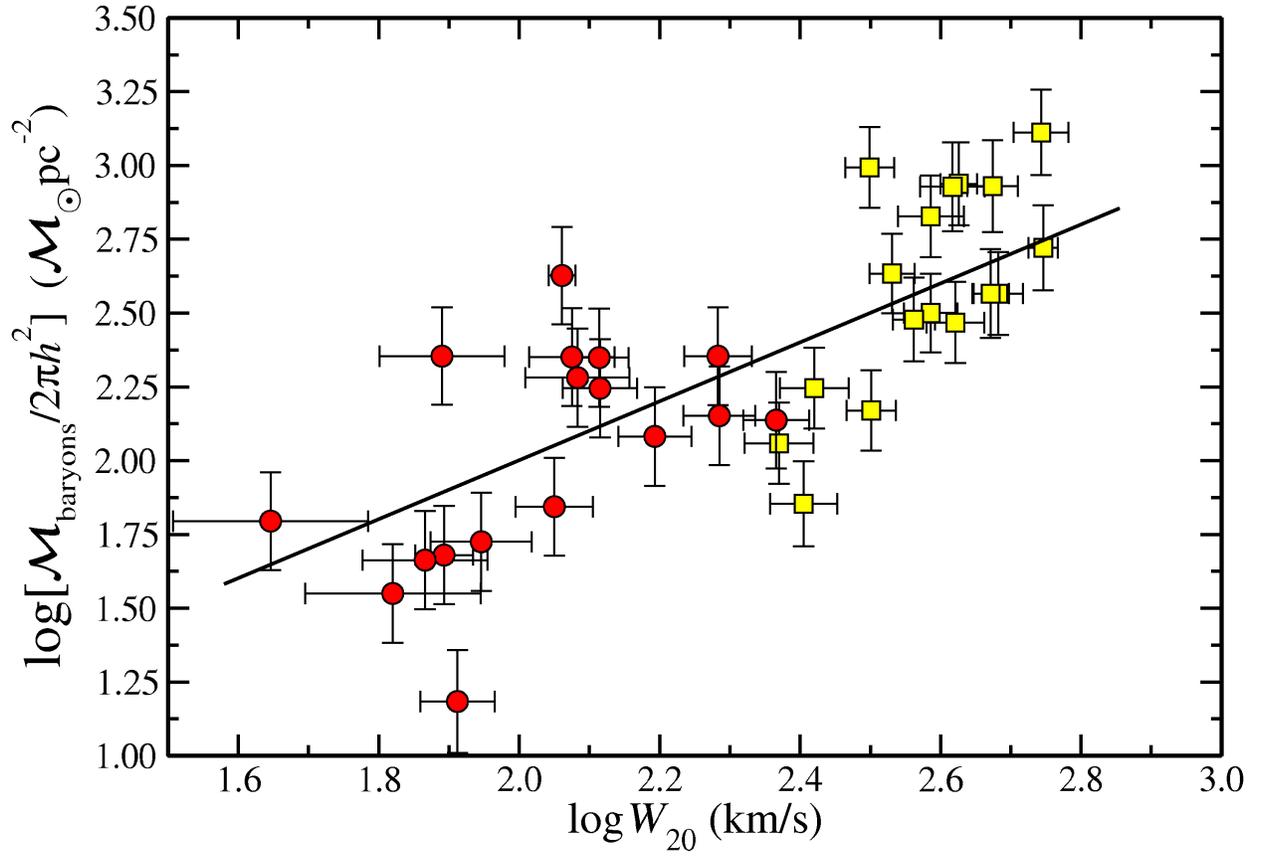}
  \caption{Mean baryon mass surface density \textit{vs} \velw for the
    two samples of disk galaxies: \textrm{H\,\scriptsize{I}} selected
    (red circles) and \cite{sak2000} (yellow squares). A line with
    slope $\sim 1$ is shown, for comparison.
     \label{denvsw20_1}}
\end{figure}

As mentioned in Sec.\,\ref{sec:results} we examine the effect that
scaling our modelled $\mathcal{M}$/$L$ values would have on the BTFR
for the \cite{sak2000} galaxies. We find that if we scale our
$\mathcal{M}$/$L$ by 1.5 and 2.0, weighted bivariate fits are produced
with reduced $\chi^2 = 0.9$ that are significantly steeper, with
slopes: $3.3 \pm 0.3$ and $3.5 \pm 0.3$, respectively. Even larger
scalings, indicative of a significantly different IMF, are unlikely to
be real given the agreement as shown in Figure \ref{f_sps} between our
modelled $\mathcal{M}$/$L$ and those of Bell \& de Jong (2001) who
argue for a universal scaled Salpeter IMF by taking maximum disk
dynamical constraints into account. However, some authors
\cite[$\eg$][]{meureretal2009}, argue for a non universal IMF and that
the upper end of the IMF varies systematically with galaxy mass. If
true, low luminosity galaxies may have less massive stars than high
luminosity galaxies and this may imply that even more stellar mass is
locked up in the dwarf galaxies than we have accounted for with our
universal Chabrier IMF. This would seem to be partial evidence against
the large $\mathcal{M}$/$L$ scalings that produce a steeper BTFR slope.

Although, some authors prefer to use the flat part of the rotation
curve as their velocity estimator, rather than $W_{20}$, choosing
galaxies for which the flat part of the rotation curve is observable
provides a consistent estimate of rotation but it is itself a strong
selection effect which most low-mass galaxies do not satisfy. Samples
of low-mass galaxies for which the flat part of the rotation curve is
observable are biased towards those for which the
\textrm{H\,\scriptsize{I}} distribution is more extended relative to
the halo scale length $r_s$: i.e. their \textrm{H\,\scriptsize{I}} is
intrinsically more extended or they have more centrally concentrated
halos with relatively smaller scale lengths $r_s$ and relatively
larger concentration parameters $c = r_{vir}/r_s$.  We can speculate
why such a bias could tilt the slope of the BTFR towards higher values
($\sim 4$) as observed: for example, it is possible that a significant
fraction of the hydrogen in these more extended low-mass galaxies is
ionized by the metagalactic UV field.  In any case, the existence of
this selection effect needs to be recognized.

For the purposes of relating dark matter and baryonic matter via the
BTFR, we should also ask whether the flat level of the rotation curve
V$_{\rm{flat}}$ is the right velocity to use ?  V$_{\rm{ flat}}$ is
unlikely to be a good estimator of the circular velocity V$_{\rm
  {vir}}$ at the virial radius. We know this from basic theory ($\eg$
of NFW models) and from the observational work of \cite{bat05} for our
own Galaxy, which indicates that V$_{\rm {flat}} >$ V$_{\rm {vir}}$.
Is V$_{\rm {flat}}$ any better for estimating V$_{\rm {vir}}$ than
V$_{\rm {max}}$ of the maximum $\textrm{H\,\scriptsize{I}}$ velocity
which determines $W_{20}$?

Even if $\mathcal{M}_{\rm baryons} \propto \rm{V}_{\rm {flat}}^{4}$,
how does that relate to the cosmological slope of the $\mathcal{M}_{\rm
baryons}$ \textit{vs} V$_{\rm{vir}}$ relation ? For example, to reconcile the
usual BTFR slope of 4 with the cosmological slope of 3, we would need
V$_{\rm {flat}} \propto$ V$_{\rm {vir}}^{3/4}$: $i.e.,\;$ a systematic
change in halo structure with V$_{\rm {vir}}$. Most lower-mass galaxies
have V$_{\rm {max}}<$ V$_{\rm {flat}}$ while most higher-mass galaxies
have V$_{\rm {max}} \approx$ V$_{\rm {flat}}$. This observed trend of
V$_{\rm {max}}$/V$_{\rm {flat}}$ could in fact make V$_{\rm {max}}$
($i.e.,\;$ via
$W_{20}$) a more consistent estimator of V$_{\rm {vir}}$ than is V$_{\rm
 {flat}}$.  Relating V$_{\rm {max}}$ or V$_{\rm {flat}}$ to the
V$_{\rm {vir}}$ remains an unsolved problem, and it is not clear which
of V$_{\rm {max}}$ or V$_{\rm {flat}}$ is better for our problem of
relating the BTFR to cosmology.

\acknowledgments SG thanks the RSAA/ATNF/IATE students and staff for
their assistance and advice as well as Stephane Courteau, Renee Kraan
Korteweg, Stacy McGaugh, James Schombert, Joe Silk, and John Tonry for
useful discussion. This research was supported by an ANU postgraduate
research scholarship, an ATNF travel scholarship, an Argentine CONICET
fellowship and has made use of the NASA/IPAC Extragalactic Database
(NED) which is operated by the Jet Propulsion Laboratory, California
Institute of Technology, under contract with the National Aeronautics
and Space Administration. IP thanks the Mexican foundation
CONACYT. Finally, we are also very grateful to the anonymous referee
whose many comments contributed to a much improved paper.




\begin{deluxetable}{lccl}
\tabletypesize{\scriptsize}
\tablewidth{0pt}
\tablecaption{The \textrm{H\,\scriptsize{I}} quantities of the \cite{sak2000} galaxies \label{tab:sakhi}}
\tablehead{
\colhead{\textbf{Galaxy }} &
\colhead{\textbf{\textrm{H\,\scriptsize{I}} Flux (Jy\,km\,s}$^\mathbf{{-1}}$\textbf{)}} &
\colhead{\textbf{log}$\mathbf{W_{20}}$\textbf{(km/s)}} & 
\colhead{\textbf{References}}}  
\startdata
M31         & $36732.2$ & $2.74\,\pm 0.03$ &$1,2,3$ \\
M33         & $13501.4$ & $2.40\,\pm 0.07$ &$1,2,3$ \\
NGC\phd925  & $328.5$ & $2.42\,\pm 0.05$   &$1,2,3,4$ \\
NGC\phd1365 & $168.1$ & $2.68\,\pm 0.04$   & $2,3,5,6,7 $\\
NGC\phd1425 & $51.7$  & $2.62\,\pm 0.04$   & $4,5,6,7$ \\
NGC\phd2090 & $125.3$ & $2.50\,\pm 0.04$   & $3,5,6$ \\
NGC\phd2403 & $1547.3$  & $2.48\,\pm 0.06$   & $1,2,3,4$ \\
NGC\phd2541 & $145.1$ & $2.37\,\pm 0.05$   & $1,2,3$ \\
NGC\phd3031 & $795.4$ & $2.72\,\pm 0.03$   &$2,3,4,8$ \\
NGC\phd3198 & $238.4$ & $2.53\,\pm 0.03$   & $1,2,3,4$ \\
NGC\phd3319 & $83.1$  & $2.41\,\pm 0.05$   &$3,9$ \\
NGC\phd3351 & $58.5$  & $2.59\,\pm 0.05$   & $1,2,3,4$ \\
NGC\phd3368 & $81.5$  & $2.67\,\pm 0.04$   & $1,3$ \\
NGC\phd3621 & $799.4$ & $2.50\,\pm 0.04$   & $2,3,4,5$ \\
NGC\phd3627 & $41.7$  & $2.63\,\pm 0.03$   & $1,3,4$ \\
NGC\phd4414 & $67.7$  & $2.74\,\pm 0.04$   & $3,10$ \\
NGC\phd4535 & $80.8$  & $2.59\,\pm 0.04$   & $3,11,12$ \\
NGC\phd4536 & $93.9$  & $2.56\,\pm 0.03$   & $3,4$ \\
NGC\phd4548 & $12.2$  & $2.62\,\pm 0.05$   & $3,11,12$ \\
NGC\phd4725 & $110.2$ & $2.67\,\pm 0.03$   & $1,2,3,4$ \\
NGC\phd7331 & $185.7$ & $2.75\,\pm 0.02$   & $1,2,3,4,13$ \\
\enddata
\tablerefs{(1) \cite{mar98}; (2) \cite{pil04}; (3) \cite{dev92}; (4) \cite{ken03}; (5) \cite{kor2004}; (6) \cite{mat96}; (7) \cite{sch01}; (8) \cite{lan03}; (9) \cite{bro94}; (10) \cite{val02}; (11) \cite{cay90b}; (12) \cite{huc89}; (13) \cite{tif90}} \\
\end{deluxetable}

\begin{deluxetable}{lccccccc}
\tabletypesize{\scriptsize}
\tablewidth{0pt}
\tablecaption{Photometrically derived quantities of the \cite{sak2000} galaxies \label{tab:dis_sak}}
\tablehead{
\colhead{\textbf{Galaxy}} &
\colhead{\textbf{dist.}} &
\colhead{\textbf{$\mathbf{\mathcal{M}}\phd$\scriptsize{gas}}} &
\colhead{\textbf{M$_{V}$}} & 
\colhead{\textbf{M$_{I}$}}  &
\colhead{\textbf{M$_{H}$}} &
\colhead{\textbf{[Fe/H]}} &
\colhead{\textbf{$h$}}}
\startdata
M31\tablenotemark{a,b,d} &$0.77 \pm      0.04$& $6.43\,(1.71)$\,E\,+09  & \nodata    & $-23.11 \pm	0.18$ &	$-24.58 \pm	0.11$ &	$-0.30 \pm0.50$             &  \nodata \\
M33\tablenotemark{a,b,d} &$0.85  \pm      0.04$& $2.87\,(0.76)$\,E\,+09  & \nodata    & $-19.84 \pm	0.18$ &	$-21.06 \pm	0.11$ &	$-0.46 \pm0.70$             &  \nodata \\
NGC\,925               &$9.29  \pm      0.34$& $9.37\,(2.44)$\,E\,+09  &$-20.10 \pm      0.16$& $-20.67 \pm	0.28$ &	$-21.91 \pm	0.10$ &	$-0.73 \pm	0.15$ & $3.84$ \\
NGC\,1365	       &$18.97 \pm     1.75$& $2.00\,(0.62)$\,E\,+10  &$-22.34 \pm	0.12$& $-23.40 \pm	0.12$ &	$-25.05 \pm	0.11$ &	$-0.32 \pm	0.20$ & $8.94$ \\
NGC\,1425	       &$23.01 \pm     0.64$& $9.05\,(2.32)$\,E\,+09  &$-21.44 \pm	0.08$& $-22.39 \pm	0.08$ &	$-23.82 \pm	0.08$ &	$-0.28 \pm	0.15$ & $5.73$ \\
NGC\,2090	       &$12.30 \pm     0.45$& $6.26\,(1.63)$\,E\,+09  &$-20.27 \pm	0.10$& $-21.30 \pm	0.11$ &	$-22.70 \pm	0.10$ &	$-0.48 \pm	0.15$ & $5.29$ \\
NGC\,2403\tablenotemark{a,b}&$3.18 \pm 0.35$& $5.16\,(1.72)$\,E\,+09  & \nodata    & $-20.38 \pm	0.28$ &	$-21.89 \pm	0.25$ &	$-0.48 \pm	        0.40$ & \nodata \\
NGC\,2541	       &$12.42 \pm     0.46$& $7.39\,(1.93)$\,E\,+09  &$-19.19 \pm	0.13$& $-19.87 \pm	0.13$ &	$-20.99 \pm	0.10$ &	$-0.78 \pm	0.15$ & $3.76$ \\
NGC\,3031\tablenotemark{a,b}&$3.63 \pm 0.13$& $3.46\,(0.90)$\,E\,+09  & \nodata    & $-22.50 \pm	0.17$ &	$-24.29 \pm	0.10$ &	$-0.53 \pm	0.15$         & \nodata \\
NGC\,3198	       &$14.45 \pm     0.40$& $1.65\,(0.42)$\,E\,+10  &$-20.78 \pm	0.08$& $-21.56 \pm	0.08$ &	$-22.97 \pm	0.08$ &	$-0.68 \pm	0.15$ & $3.36$ \\
NGC\,3319	       &$14.32 \pm     0.79$& $5.63\,(1.54)$\,E\,+09  &$-19.72 \pm	0.14$& $-20.33 \pm	0.14$ &	$-21.19 \pm	0.13$ &	$-0.90 \pm	0.15$ & $4.66$ \\
NGC\,3351	       &$10.05 \pm     0.37$& $1.95\,(0.51)$\,E\,+09  &$-20.61 \pm	0.09$& $-21.64 \pm	0.09$ &	$-23.31 \pm	0.10$ &	$-0.04 \pm	0.20$ & $2.91$ \\
NGC\,3368	       &$10.96 \pm     0.51$& $3.24\,(0.86)$\,E\,+09  &$-21.31 \pm	0.12$& $-22.31 \pm	0.11$ &	$-24.16 \pm	0.11$ &	$-0.08 \pm	0.10$ & $3.34$ \\
NGC\,3621	       &$6.70 \pm      0.34$& $1.19\,(0.32)$\,E\,+10  &$-20.16 \pm	0.12$& $-21.06 \pm	0.12$ &	$-22.59 \pm	0.12$ &	$-0.53 \pm	0.15$ & $2.10$ \\
NGC\,3627	       &$10.28 \pm     0.81$& $1.46\,(0.43)$\,E\,+09  &$-21.79 \pm	0.18$& $-22.66 \pm	0.18$ &	$-24.25 \pm	0.18$ &	$-0.03 \pm	0.20$ & $3.21$ \\
NGC\,4414	       &$19.14 \pm     0.88$& $8.20\,(2.19)$\,E\,+09  &$-21.60 \pm	0.11$& $-22.61 \pm	0.11$ &	$-24.40 \pm	0.11$ &	$-0.08 \pm	0.15$ & $3.28$ \\
NGC\,4535	       &$16.60 \pm     0.54$& $7.35\,(1.90)$\,E\,+09  &$-21.42 \pm	0.09$& $-22.28 \pm	0.08$ &	$-23.47 \pm	0.09$ &	$-0.08 \pm	0.15$ & $4.31$ \\
NGC\,4536	       &$15.49 \pm     0.57$& $7.44\,(1.94)$\,E\,+09  &$-21.05 \pm	0.10$& $-21.95 \pm	0.13$ &	$-23.54 \pm	0.09$ &	$-0.43 \pm	0.20$ & $4.58$ \\
NGC\,4548\tablenotemark{c}&$16.14 \pm  1.71$& $1.05\,(0.34)$\,E\,+09  &$-21.11 \pm	0.23$& $-22.17 \pm	0.23$ &	\nodata &	$ 0.06 \pm	0.15$         & $3.09$ \\
NGC\,4725	       &$13.00 \pm     0.48$& $6.16\,(1.60)$\,E\,+09  &$-21.91 \pm	0.09$& $-22.77 \pm	0.09$ &	$-24.40 \pm	0.10$ &	$-0.36 \pm	0.15$ & $5.39$ \\
NGC\,7331	       &$15.07 \pm     0.69$& $1.39\,(0.37)$\,E\,+10  &$-22.48 \pm	0.11$& $-23.38 \pm	0.11$ &	$-25.41 \pm	0.11$ &	$-0.61 \pm	0.15$ & $5.74$ \\
\enddata
\tablecomments{Column (1) is the galaxy name; Column (2) is the
  adopted distance (in Mpc); Column (3) is the derived gas mass (in $\mathcal{M}_\odot$); Column
  (4) is the absolute $V$-band magnitude; Column (5) is the absolute
  $I$-band magnitude; Column (6) is the absolute $H$-band magnitude; Column (7) is adopted metallicity; Column (8) is the calculated scalelength (in kpc) determined from the $V$-band surface brightness profiles from \cite{mac2000}\\Unless indicated all [Fe/H] values calculated from \cite{fer00}}
\tablenotetext{a}{M$_{V}$ unavailable in \cite{sak2000}}
\tablenotetext{b}{Surface brightness data unavailable in \cite{mac2000}}
\tablenotetext{c}{Unreliable M$_{H}$ in \cite{sak2000}}
\tablenotetext{d}{[Fe/H] from \cite{allenandshanks2004}}
\textbf{For all tables when expressed in scientific notation, errors are in braces}\\ \textbf{NOTE:} $6.43\,(1.71)$\,E\,+09 denotes $6.43$ x $10^9 \pm 1.71$ x $10^9$

\end{deluxetable}

\begin{deluxetable}{lcccccc}
\tablewidth{0pt}
\tabletypesize{\scriptsize}
\tablecaption{Evolutionary stellar population synthesis modelling for the Sakai galaxies from ($V-I$) constrained simulations \label{tab:dis_mlt_sak_vmi}}
\tablehead{
\colhead{\textbf{Galaxy}} &
\colhead{$(\mathcal{M}/L_V)$}&
\colhead{$(\mathcal{M}/L_I)$} & 
\colhead{$\mathcal{M}_{\textrm{stars}}:V$}  &
\colhead{$\mathcal{M}_{\textrm{stars}}:I$} &
\colhead{$\mathcal{M}_{\textrm{total}}:V$} & 
\colhead{$\mathcal{M}_{\textrm{total}}:I$}}
\startdata
NGC\,925   & $0.82 \pm	   0.29$ &	$0.74 \pm  0.20$ & $8.33\,(3.13)$\,E\,+09 & $6.48\,(2.42)$\,E\,+09 &$1.77\,(0.41)$\,E\,+10 & $1.58\,(0.35)$\,E\,+10 \\
NGC\,1365  & $1.82 \pm       0.26$ &	$1.34 \pm  0.15$ & $1.45\,(0.35)$\,E\,+11 & $1.44\,(0.32)$\,E\,+11 &$1.65\,(0.38)$\,E\,+11 & $1.64\,(0.36)$\,E\,+11 \\
NGC\,1425  & $1.45 \pm       0.47$ &	$1.14 \pm  0.26$ & $5.04\,(1.65)$\,E\,+10 & $4.82\,(1.17)$\,E\,+10 &$5.94\,(1.68)$\,E\,+10 & $5.73\,(1.20)$\,E\,+10 \\
NGC\,2090  & $1.57 \pm       0.37$ &	$1.20 \pm  0.21$ & $1.85\,(0.52)$\,E\,+10 & $1.87\,(0.38)$\,E\,+10 &$2.47\,(0.56)$\,E\,+10 & $2.49\,(0.43)$\,E\,+10 \\
NGC\,2403\tablenotemark{a}  & \nodata & \nodata & \nodata & \nodata & \nodata & \nodata \\
NGC\,2541  & $0.63 \pm       0.09$ &	$0.62 \pm  0.07$ & $2.76\,(0.48)$\,E\,+09 & $2.57\,(0.41)$\,E\,+09 &$1.01\,(0.20)$\,E\,+10 & $9.96\,(2.02)$\,E\,+09 \\
NGC\,3031\tablenotemark{a}  & \nodata & \nodata & \nodata & \nodata & \nodata & \nodata  \\                 
NGC\,3198  & $0.72 \pm       0.12$ &	$0.68 \pm  0.09$ & $1.36\,(0.25)$\,E\,+10 & $1.34\,(0.21)$\,E\,+10 &$3.01\,(0.50)$\,E\,+10 & $2.98\,(0.48)$\,E\,+10 \\
NGC\,3319\tablenotemark{b}  & \nodata & \nodata & \nodata & \nodata & \nodata & \nodata  \\
NGC\,3351  & $1.72 \pm       0.59$ &	$1.29 \pm  0.33$ & $2.77\,(0.98)$\,E\,+10 & $2.75\,(0.73)$\,E\,+10 &$2.97\,(0.99)$\,E\,+10&$2.95\,(0.74)$\,E\,+10\\
NGC\,3368  & $1.57 \pm       0.27$ &	$1.22 \pm  0.17$ & $4.83\,(0.97)$\,E\,+10 & $4.81\,(0.82)$\,E\,+10 &$5.15\,(0.99)$\,E\,+10&$5.14\,(0.84)$\,E\,+10\\
NGC\,3621  & $1.16 \pm       0.34$ &	$0.96 \pm  0.21$ & $1.23\,(0.38$)\,E\,+10 & $1.20\,(0.29)$\,E\,+10 &$2.42\,(0.53)$\,E\,+10&$2.38\,(0.46)$\,E\,+10\\
NGC\,3627  & $1.01 \pm       0.19$ &	$0.88 \pm  0.13$ & $4.84\,(1.20)$\,E\,+10 & $4.78\,(1.06)$\,E\,+10 &$4.99\,(1.22)$\,E\,+10&$4.93\,(1.07)$\,E\,+10\\
NGC\,4414  & $1.62 \pm       0.28$ &	$1.24 \pm  0.15$ & $6.49\,(1.28)$\,E\,+10 & $6.47\,(1.01)$\,E\,+10 &$7.31\,(1.34)$\,E\,+10&$7.29\,(1.08)$\,E\,+10\\
NGC\,4535  & $0.97 \pm       0.17$ &	$0.85 \pm  0.12$ & $3.32\,(0.62)$\,E\,+10 & $3.28\,(0.50)$\,E\,+10 &$4.05\,(0.66)$\,E\,+10&$4.01\,(0.55)$\,E\,+10\\
NGC\,4536  & $1.11 \pm       0.20$ &	$0.94 \pm  0.12$ & $2.70\,(0.54)$\,E\,+10 & $2.67\,(0.47)$\,E\,+10 &$3.44\,(0.59)$\,E\,+10&$3.42\,(0.53)$\,E\,+10\\
NGC\,4548  & $1.94 \pm       0.28$ &	$1.43 \pm  0.15$ & $4.97\,(1.29)$\,E\,+10 & $4.95\,(1.19)$\,E\,+10 &$5.08\,(1.31)$\,E\,+10&$5.06\,(1.21)$\,E\,+10\\
NGC\,4725  & $0.95 \pm       0.16$ &	$0.84 \pm  0.11$ & $5.08\,(0.96)$\,E\,+10 & $5.06\,(0.79)$\,E\,+10 &$5.70\,(0.99)$\,E\,+10&$5.68\,(0.83)$\,E\,+10\\
NGC\,7331  & $1.05 \pm       0.24$ &	$0.90 \pm  0.15$ & $9.50\,(2.35)$\,E\,+10 & $9.51\,(1.85)$\,E\,+10 &$1.09\,(0.24)$\,E\,+11&$1.09\,(0.19)$\,E\,+11\\
\enddata
\tablecomments{Column (1) is the galaxy name; Column (2) is the $V$-band $\mathcal{M}$/L (in solar units); Column
  (3) is the $I$-band $\mathcal{M}$/L (in solar units); Column (4) is the stellar mass (in solar units) derived from the $V$-band data; Column (5) is the stellar mass (in solar units) derived from the $I$-band data; Column (6) is the total mass (in solar units) derived using the $V$-band data; Column (7) is the total mass (in solar units) derived using the $I$-band data}
\tablenotetext{a}{Not modeled: no $V$-band magnitude available from \cite{sak2000}}
\tablenotetext{b}{Excluded: the model age is less than 8 Gyr}
\end{deluxetable}

\begin{deluxetable}{lcccccc}
\tablewidth{0pt}
\tabletypesize{\scriptsize}
\tablecaption{Evolutionary stellar population synthesis modelling for the Sakai galaxies from ($V-H$) constrained simulations \label{tab:dis_mlt_sak_vmh}}
\tablehead{
\colhead{\textbf{Galaxy}} &
\colhead{$(\mathcal{M}/L_V)$}&
\colhead{$(\mathcal{M}/L_H)$}& 
\colhead{$\mathcal{M}_{\textrm{stars}}:V$}  &
\colhead{$\mathcal{M}_{\textrm{stars}}:H$} &
\colhead{$\mathcal{M}_{\textrm{total}}:V$} & 
\colhead{$\mathcal{M}_{\textrm{total}}:H$}}
\startdata
NGC\,925   & $0.69 \pm       0.12 $ &	$ 0.47 \pm 0.07$ & $6.95\,(1.58)$\,E\,+09 & $6.52\,(1.17)$\,E\,+09 & $1.63\,(0.30)$\,E\,+10&$1.59\,(0.28)$\,E\,+10\\
NGC\,1365  & $2.51 \pm       0.49 $ &	$ 0.88 \pm 0.11$ & $1.99\,(0.55)$\,E\,+11 & $2.19\,(0.49)$\,E\,+11 & $2.19\,(0.58)$\,E\,+11&$2.39\,(0.52)$\,E\,+11\\
NGC\,1425  & $1.47 \pm       0.31 $ &	$ 0.70 \pm 0.12$ & $5.08\,(1.13)$\,E\,+10 & $5.56\,(1.07)$\,E\,+10 & $5.99\,(1.16)$\,E\,+10&$6.46\,(1.11)$\,E\,+10\\
NGC\,2090  & $1.74 \pm       0.36 $ &	$ 0.74 \pm 0.10$ & $2.06\,(0.53)$\,E\,+10 & $2.09\,(0.35)$\,E\,+10 & $2.69\,(0.57)$\,E\,+10&$2.72\,(0.40)$\,E\,+10\\
NGC\,2403\tablenotemark{a}  & \nodata & \nodata & \nodata & \nodata & \nodata & \nodata \\
NGC\,2541  & $0.69 \pm       0.18 $ &	$ 0.47 \pm 0.09$ & $3.03\,(0.84)$\,E\,+09 & $2.77\,(0.57)$\,E\,+09 & $1.04\,(0.22)$\,E\,+10&$1.02\,(0.21)$\,E\,+10\\
NGC\,3031\tablenotemark{a}  & \nodata & \nodata & \nodata & \nodata & \nodata & \nodata \\
NGC\,3198  & $0.67 \pm       0.12 $ &	$ 0.46 \pm 0.07$ & $1.26\,(0.25)$\,E\,+10 & $1.69\,(0.28)$\,E\,+10 & $2.90\,(0.50)$\,E\,+10&$3.33\,(0.52)$\,E\,+10\\
NGC\,3319  & $0.72 \pm       0.15 $ &	$ 0.49 \pm 0.08$ & $5.10\,(1.23)$\,E\,+09 & $3.44\,(0.70)$\,E\,+09 & $1.07\,(0.21)$\,E\,+10&$9.07\,(1.82)$\,E\,+09\\
NGC\,3351  & $2.22 \pm       0.45 $ &	$ 0.79 \pm 0.11$ & $3.59\,(0.79)$\,E\,+10 & $3.95\,(0.63)$\,E\,+10 & $3.79\,(0.79)$\,E\,+10&$4.15\,(0.64)$\,E\,+10\\
NGC\,3368  & $2.80 \pm       0.59 $ &	$ 0.87 \pm 0.13$ & $8.61\,(2.02)$\,E\,+10 & $0.95\,(0.17)$\,E\,+11 & $8.94\,(2.03)$\,E\,+10&$0.98\,(0.17)$\,E\,+11\\
NGC\,3621  & $1.82 \pm       0.36 $ &	$ 0.76 \pm 0.11$ & $1.94\,(0.43)$\,E\,+10 & $1.96\,(0.35)$\,E\,+10 & $3.13\,(0.58)$\,E\,+10&$3.15\,(0.52)$\,E\,+10\\
NGC\,3627  & $1.53 \pm       0.34 $ &	$ 0.66 \pm 0.14$ & $7.34\,(2.01)$\,E\,+10 & $7.88\,(2.06)$\,E\,+10 & $7.49\,(2.03)$\,E\,+10&$8.03\,(2.07)$\,E\,+10\\
NGC\,4414  & $2.65 \pm       0.41 $ &	$ 0.86 \pm 0.07$ & $1.07\,(0.20)$\,E\,+11 & $1.16\,(0.16)$\,E\,+11 & $1.15\,(0.20)$\,E\,+11&$1.25\,(0.16)$\,E\,+11\\
NGC\,4535  & $0.83 \pm       0.14 $ &	$ 0.46 \pm 0.06$ & $2.83\,(0.54)$\,E\,+10 & $2.64\,(0.38)$\,E\,+10 & $3.56\,(0.58)$\,E\,+10&$3.38\,(0.45)$\,E\,+10\\
NGC\,4536  & $1.91 \pm       0.37 $ &	$ 0.76 \pm 0.11$ & $4.63\,(0.98)$\,E\,+10 & $4.69\,(0.80)$\,E\,+10 & $5.38\,(1.02)$\,E\,+10&$5.43\,(0.84)$\,E\,+10\\
NGC\,4548\tablenotemark{b}  & \nodata & \nodata & \nodata & \nodata & \nodata & \nodata \\
NGC\,4725  & $1.87 \pm       0.32 $ &	$ 0.74 \pm 0.10$ & $9.99\,(1.89)$\,E\,+10 & $1.00\,(0.16)$\,E\,+11 & $1.06\,(0.19)$\,E\,+11&$1.06\,(0.16)$\,E\,+11\\
NGC\,7331\tablenotemark{c}  & \nodata & \nodata & \nodata & \nodata & \nodata & \nodata \\
\enddata
\tablecomments{Column (1) is the galaxy name; Column (2) is the $V$-band $\mathcal{M}$/L (in solar units); Column
  (3) is the $H$-band $\mathcal{M}$/L (in solar units); Column (4) is the stellar mass (in solar units) using the $V$-band data; Column (5) is the stellar mass (in solar units) using the $H$-band data; Column (6) is the total mass (in solar units) derived using the $V$-band data; Column (7) is the total mass (in solar units) derived using the $H$-band data }
\tablenotetext{a}{Not modeled: no $V$-band magnitude available from \cite{sak2000}}
\tablenotetext{b}{Not modeled: no $H$-band magnitude is available from \cite{sak2000}}
\tablenotetext{c}{Unable to model: no model color exists as red as the observed given the metallicity}
\end{deluxetable}

\begin{deluxetable}{lcccccc}
\tablewidth{0pt}
\tabletypesize{\scriptsize}
\tablecaption{Derived baryon masses of the Sakai galaxies \label{tab:sak_mas_den}}
\tablehead{
\colhead{\textbf{Galaxy}} &
\colhead{$\mathbf{\log W_{20}}$} &
\colhead{$\mathbf{\mathcal{F}}_{gas}$} &
\colhead{$\mathbf{\mathcal{M}_{stars}}$} & 
\colhead{$\mathbf{\mathcal{M}_{total}}$}  &
\colhead{$\mathbf{\log\Sigma_{stars}}$} & 
\colhead{$\mathbf{\log\Sigma_{total}}$}} 
\startdata
NGC\,925                   &$2.42 \pm 0.05$&$1.38\pm 0.40$ & $6.77\,(0.85)$\,E\,+09&     $1.63\,(0.16)$\,E\,+10&$1.86$&$2.25$\\    
NGC\,1365 	         &$2.68 \pm 0.04$&$0.12\pm 0.04$ & $1.64\,(0.20)$\,E\,+11&     $1.85\,(0.22)$\,E\,+11&$2.51$&$2.57$\\   
NGC\,1425 	         &$2.62 \pm 0.04$&$0.18\pm 0.05$ & $5.16\,(0.60)$\,E\,+10&     $6.06\,(0.62)$\,E\,+10&$2.40$&$2.47$\\   
NGC\,2090 	         &$2.50 \pm 0.04$&$0.32\pm 0.09$ & $1.98\,(0.21)$\,E\,+10&     $2.60\,(0.24)$\,E\,+10&$2.05$&$2.17$\\   
NGC\,2403\tablenotemark{a} &\nodata&\nodata&\nodata        &\nodata&\nodata&\nodata\\    			       	                 
NGC\,2541 	         &$2.37 \pm 0.05$&$2.73\pm 0.75$ & $2.71\,(0.26)$\,E\,+09&     $1.02\,(0.10)$\,E\,+10&$1.48$&$2.06$\\   
NGC\,3031\tablenotemark{a} &\nodata&\nodata&\nodata        &\nodata&\nodata&\nodata\\	    		       	                       
NGC\,3198 	         &$2.53 \pm 0.03$&$1.18\pm 0.32$&$1.39\,(0.12)$\,E\,+10&       $3.05\,(0.25)$\,E\,+10&$2.29$&$2.63$\\   
NGC\,3319 	         &$2.41 \pm 0.05$&$1.46\pm 0.89$&$3.84\,(0.61)$\,E\,+09&       $9.77\,(1.39)$\,E\,+09&$1.45$&$1.85$\\   
NGC\,3351 	         &$2.59 \pm 0.05$&$0.06\pm 0.02$&$3.38\,(0.38)$\,E\,+10&       $3.57\,(0.38)$\,E\,+10&$2.80$&$2.83$\\   
NGC\,3368 	         &$2.67 \pm 0.04$&$0.06\pm 0.02$&$5.61\,(0.97)$\,E\,+10&       $5.95\,(0.98)$\,E\,+10&$2.91$&$2.93$\\   
NGC\,3621 	         &$2.50 \pm 0.04$&$0.78\pm 0.23$&$1.52\,(0.21)$\,E\,+10&       $2.73\,(0.26)$\,E\,+10&$2.74$&$2.99$\\   
NGC\,3627 	         &$2.63 \pm 0.03$&$0.03\pm 0.01$&$5.46\,(0.70)$\,E\,+10&       $5.61\,(0.71)$\,E\,+10&$2.93$&$2.94$\\   
NGC\,4414 	         &$2.74 \pm 0.04$&$0.10\pm 0.03$&$7.90\,(1.27)$\,E\,+10&       $8.76\,(1.28)$\,E\,+10&$3.07$&$3.11$\\   
NGC\,4535 	         &$2.59 \pm 0.04$&$0.25\pm 0.07$&$2.93\,(0.24)$\,E\,+10&       $3.69\,(0.27)$\,E\,+10&$2.40$&$2.50$\\   
NGC\,4536 	         &$2.56 \pm 0.03$&$0.23\pm 0.07$&$3.17\,(0.49)$\,E\,+10&       $3.96\,(0.51)$\,E\,+10&$2.38$&$2.48$\\   
NGC\,4548                  &$2.62 \pm 0.05$&$0.02\pm 0.01$&$4.96\,(0.88)$\,E\,+10&       $5.07\,(0.89)$\,E\,+10&$2.92$&$2.93$\\   
NGC\,4725 	         &$2.67 \pm 0.03$&$0.10\pm 0.03$&$6.05\,(1.14)$\,E\,+10&       $6.71\,(1.16)$\,E\,+10&$2.52$&$2.57$\\   
NGC\,7331                  &$2.75 \pm 0.02$&$0.15\pm 0.05$&$9.51\,(1.45)$\,E\,+10&       $1.09\,(0.15)$\,E\,+11&$2.66$&$2.72$\\   
\enddata
\tablenotetext{a}{Not modeled: no $V$-band magnitude available from \cite{sak2000}}
\tablecomments{Column (1) is the galaxy name; Column (2) is the inclination corrected $W_{20}$ (in km\,s$^{-1}$); Column (3) is the
  gas--to--stars ratio ($\mathbf{\mathcal{M}_{gas}}$/$\mathbf{\mathcal{M}_{stars}}$); Column (4) is the stellar mass (in solar units); Column
  (5) is the gas plus stellar mass (in solar units); Column (6) is the mean
  stellar mass surface density (in $\mathcal{M}$$_{\odot}$\,pc$^{-2}$); Column (7)
  is the mean baryon mass surface density (in $\mathcal{M}$$_{\odot}$\,pc$^{-2}$)}
\end{deluxetable}

\begin{deluxetable}{lcccc}
\tabletypesize{\scriptsize}
\tablewidth{0pt}
\tablecaption{Optical and near-IR surface photometry for the \textrm{H\,\scriptsize{I}} sample\label{tab:hiphot1}}
\tablehead{
\colhead{\textbf{Galaxy}} &
\colhead{$\mathbf{\mu_{\it{H}}}$} &
\colhead{$\mathbf{\it{h_{H}}}$} & 
\colhead{$\mathbf{\mu_{\it{V}}}$} &
\colhead{$\mathbf{\it{h_{V}}}$}}  
\startdata
HIPASS\phd J1112-86	& $19.33 $ & $9.99 $ &	$20.60 $ & $5.98 $  \\
HIPASS\phd J0554-71	& $19.00 $ & $5.43 $ &	$21.76 $ & $5.65 $  \\
HIPASS\phd J1934-67	& $19.02 $ & $11.26$ &	$20.86 $ & $10.43$  \\
AM\phd 0433-654	        & $21.48 $ & $18.54$ &	$23.18 $ & $13.40 $  \\
IC\phd 5028		& $20.07 $ & $13.78$ &	$22.18 $ & $16.55$  \\
IC\phd 5008                & $18.51 $ & $6.27 $ &   $21.23 $ & $10.09$  \\ 
ESO\phd 383-\phd G\phd 092	& $17.33 $ & $4.53 $ &	$19.12 $ & $4.66$  \\
ESO\phd 318-\phd G\phd 013	& $19.96 $ & $28.71$ &	$21.32 $ & $21.22$  \\
HIPASS\phd J1801-72\tablenotemark{a}	& $16.28 $ & $3.25 $ &	$18.87 $ & $3.94$  \\
ESO\phd 148-\phd G\phd 006	& $19.58 $ & $12.03$ &	$21.66 $ & $12.36$  \\
ESO\phd 084-\phd G\phd 040	& $19.07 $ & $11.53$ &	$20.59 $ & $10.77$  \\
HIPASS\phd J1424-16b	& $19.66 $ & $15.12$ &	$21.98 $ & $14.35$  \\
HIPASS\phd J0736-74	& $18.60 $ & $6.58 $ &	$20.04 $ & $5.85 $  \\
ESO\phd 085-\phd G\phd 088	& $20.72 $ & $49.03$ &	$22.57 $ & $38.57$  \\
SGC\phd 0454.2-6138	& $20.30 $ & $11.59$ &	$22.34 $ & $13.51$  \\
ESO\phd 052-\phd G\phd 010	& $19.60 $ & $8.14 $ &	$21.21 $ & $7.77 $  \\
ESO\phd 321-\phd G\phd 014	& $20.49 $ & $25.72$ &	$20.97 $ & $11.81$  \\
HIPASS\phd J0653-73	& $20.43 $ & $12.67$ &	$22.73 $ & $12.51$  \\
ESO\phd 140-\phd G\phd 019	& $20.68 $ & $14.17$ &	$23.07 $ & $18.78$  \\
HIPASS\phd J0039-76        & $18.07 $ & $5.40 $ &	$20.23 $ & $6.18 $  \\
\enddata
\tablecomments{Column (1) is the galaxy name; Column (2) is 
$H$-band disk central surface brightness (in mag arcsec$^{-2}$); Column
  (3) is the $H$-band scale length (in arcsec); Column (4) is the
$V$-band disk central surface brightness (in mag arcsec$^{-2}$); Column (5) is the $V$-band scalelength (in arcsec)}

\tablenotetext{a} {This galaxy appears to have a double exponential disk in the $V$-band}
\end{deluxetable}

\begin{deluxetable}{lcccccc}
\tabletypesize{\scriptsize} \tablewidth{0pt} \tablecaption{Photometric quantities of the \textrm{H\,\scriptsize{I}} selected
  sample \label{tab:hiphot_obs}} \tablehead{ \colhead{\textbf{Galaxy}}
  & \colhead{\textbf{Incl.}} & \colhead{\textbf{\textit{V}}} &
  \colhead{\textbf{\textit{H}}} & \colhead{\textbf{\textit{V-H}}}
  & \colhead{\textbf{\textit{V-H}}} &
  \colhead{\textbf{\textit{V-H}}}} \startdata HIPASS\phd J1112-86 &
$34.2$ & $15.19 \pm 0.07$ & $12.77 \pm 0.11$ & $1.97$ & $2.06$ & $2.02
\pm 0.08$ \\ HIPASS\phd J0554-71 & $37.4$ & $16.42 \pm 0.09$ & $13.98
\pm 0.13$ & $2.77$ & $2.85$ & $2.81 \pm 0.11$ \\ HIPASS\phd J1934-67 &
$53.3$ & $14.42 \pm 0.06$ & $12.40 \pm 0.06$ & $1.99$ & $1.99$ & $1.99
\pm 0.07$ \\ AM\phd 0433-654 & $42.9$ & $15.97 \pm 0.07$ & $13.43 \pm
0.10$ & $2.18$ & $2.24$ & $2.21 \pm 0.03$ \\ IC\phd 5028 & $54.6$ &
$14.71 \pm 0.08$ & $13.01 \pm 0.09$ & $1.85$ & $1.87$ & $1.86 \pm
0.10$ \\ IC\.5008 & $72.7$ & $15.05 \pm 0.05$ & $13.35 \pm 0.07$ &
$2.15$ & $2.13$ & $2.14 \pm 0.06$ \\ ESO\phd 383-\phd G\phd 092 &
$44.4$ & $14.42 \pm 0.03$ & $12.79 \pm 0.05$ & $1.60$ & $1.60$ & $1.60
\pm 0.06$ \\ ESO\phd 318-\phd G\phd 013 & $71.3$ & $14.16 \pm 0.09$ &
$12.32 \pm 0.12$ & $1.51$ & $1.55$ & $1.53 \pm 0.06$ \\ HIPASS\phd
J1801-72 & $49.5$ & $14.37 \pm 0.02$ & $12.16 \pm 0.05$ & $2.41$ &
$2.45$ & $2.43 \pm 0.11$ \\ ESO\phd 148-\phd G\phd 006 & $64.6$ &
$15.02 \pm 0.08$ & $12.99 \pm 0.09$ & $2.13$ & $2.13$ & $2.13 \pm
0.20$ \\ ESO\phd 084-\phd G\phd 040 & $48.1$ & $14.38 \pm 0.21$ &
$12.49 \pm 0.06$ & $1.65$ & $1.72$ & $1.69 \pm 0.09$ \\ HIPASS\phd
J1424-16b & $47.0$ & $14.24 \pm 0.19$ & $12.28 \pm 0.10$ & $2.39$ &
$2.39$ & $2.39 \pm 0.09$ \\ HIPASS\phd J0736-74 & $51.8$ & $15.02 \pm
0.06$ & $13.26 \pm 0.06$ & $1.56$ & $1.58$ & $1.57 \pm 0.19$
\\ ESO\phd 085-\phd G\phd 088 & $54.0$ & $13.89 \pm 0.19$ & $11.69 \pm
0.26$ & $1.95$ & $2.08$ & $2.02 \pm 0.07$ \\ SGC\phd 0454.2-6138 &
$56.2$ & $15.81 \pm 0.12$ & $13.99 \pm 0.11$ & $1.78$ & $1.78$ & $1.78
\pm 0.07$ \\ ESO\phd 052-\phd G\phd 010 & $49.5$ & $15.37 \pm 0.04$ &
$13.72 \pm 0.06$ & $1.70$ & $1.70$ & $1.70 \pm 0.09$ \\ ESO\phd
321-\phd G\phd 014 & $68.0$ & $14.96 \pm 0.06$ & $12.65 \pm 0.13$ &
$1.84$ & $1.83$ & $1.84 \pm 0.08$ \\ HIPASS\phd J0653-73 & $59.8$ &
$16.39 \pm 0.10$ & $13.87 \pm 0.10$ & $2.24$ & $2.28$ & $2.26 \pm
0.06$ \\ ESO\phd 140-\phd G\phd 019 & $56.9$ & $15.77 \pm 0.16$ &
$13.83 \pm 0.13$ & $2.10$ & $2.14$ & $2.12 \pm 0.08$ \\ HIPASS\phd
J0039-76 & $53.9$ & $14.67 \pm 0.05$ & $12.86 \pm 0.07$ & $1.83$ &
$1.83$ & $1.83 \pm 0.08$ \\ \enddata \tablecomments{Column (1) is the
  galaxy name; Column (2) is the inclination (in degrees) derived with
  a $q$ of 0.2; Column (3) is the $V$-band magnitude; Column (4) is
  the $H$-band magnitude; Column (5) is the color at the $V$-band
  half-light radius; Column (6) is the color at the $H$-band
  half-light radius; Column (7) is the adopted color from the mean of Columns (5) and (6)}
\end{deluxetable}

\begin{deluxetable}{lcccc}
\tabletypesize{\scriptsize} \tablewidth{0pt}
\tablecaption{\textrm{H\,\scriptsize{I}} quantities from our
  narrow-band observations of the \textrm{H\,\scriptsize{I}}
  sample \label{tab:pksnarrow}} \tablehead{ \colhead{\textbf{Gal.}}  &
  \colhead{\textbf{\textrm{H\,\scriptsize{I}} Flux
      (Jy\,km\,s}$^\mathbf{{-1}}$\textbf{)}} &
  \colhead{$\mathbf{v_{hel.}}$\textbf{(km\,s$^{\mathbf{-1}}$)}} &
  \colhead{\textbf{log$W\mathbf{_{20}}$\textbf{(km\,s$^{\mathbf{-1}}$)}}}
  &
  \colhead{\textbf{log$W\mathbf{_{20}}$\textbf{(km\,s$^{\mathbf{-1}}$)}}}}
\startdata HIPASS\phd J1112-86 & $3.5$ & $2187$ & $1.86$ & $2.11 \pm
0.04$ \\ HIPASS\phd J0554-71 &$3.6$ & $1482$ & $1.90$ & $2.11 \pm 0.04$
\\ HIPASS\phd J1934-67 &$3.6$ & $4123$ & $2.27$ & $2.37 \pm 0.05$
\\ AM\phd 0433-654 & $1.7$ & $1229$ & $1.65$ & $1.82 \pm 0.12$ \\ IC\phd
5028 & $9.7$ & $1619$ & $2.10$ & $2.19 \pm 0.05$ \\ IC\phd 5008 & $4.3$
& $3714$ & $2.26$ & $2.28 \pm 0.05$ \\ ESO\phd 383-\phd G\phd 092 &
$5.9$ &$1410$ & $1.86$ & $2.02 \pm 0.07$ \\ ESO\phd 318-\phd G\phd 013 &
$9.5$ & $714$ & $1.87$ & $1.89 \pm 0.04$ \\ HIPASS\phd J1801-72 & $6.0$
& $3284$ & $2.35$ & $2.47 \pm 0.02$ \\ ESO\phd 148-\phd G\phd 006
&$6.5$ & $3167$ & $2.24$ & $2.28 \pm 0.05$ \\ ESO\phd 084-\phd G\phd
040 &$6.6$ & $1235$ & $1.95$ & $2.08 \pm 0.07$ \\ HIPASS\phd J1424-16b &
$13.1$ & $1487$ & $1.93$ & $2.08 \pm 0.06$ \\ HIPASS\phd J0736-74 &
$2.2$ & $1148$ & $1.79$ & $1.89 \pm 0.09$ \\ ESO\phd 085-\phd G\phd 088
& $4.4$ & $1171$ & $1.81$ & $1.91 \pm 0.05$ \\ SGC\phd 0454.2-6138 &
$2.1$ & $972$ & $1.79$ & $1.87 \pm 0.09$ \\ ESO\phd 052-\phd G\phd 010 &
$3.8$ & $1387$ & $2.00$ & $2.11 \pm 0.05$ \\ ESO\phd 321-\phd G\phd 014
& $5.3$ & $612$ & $1.61$ & $1.65 \pm 0.14$ \\ HIPASS\phd J0653-73 &
$3.5$ & $1205$ & $1.99$ & $2.05 \pm 0.05$ \\ ESO\phd 140-\phd G\phd 019
& $4.2$ & $954$ & $1.87$ & $1.95 \pm 0.07$ \\ HIPASS\phd J0039-76 &
$3.8$ & $1754$ & $1.97$ & $2.06 \pm 0.02$ \\ \enddata
\tablecomments{Column (1) is the galaxy name; Column (2) is the
  integrated \textrm{H\,\scriptsize{I}} flux; Column (3) is the
  measured heliocentric velocity; Column (4) is the observed \velw
  measurement (not corrected for inclination); Column (5) is the \velw
  measurement (corrected for inclination)\\ The error in the
  \textrm{H\,\scriptsize{I}} flux values are 15\% determined by
  comparing our values with the independent
  \textrm{H\,\scriptsize{I}}PASS values.  v$_{\mathrm{helio.}}$ errors
  are $\simeq 5$\kms and the errors of our \velw corrected values are
  between 5 and 20\kms}
\end{deluxetable}

\begin{deluxetable}{lcccccc}
\tabletypesize{\scriptsize}
\tablewidth{0pt}
\tablecaption{Derived quantities of the \textrm{H\,\scriptsize{I}} selected sample \label{tab:dis_hipass}}
\tablehead{
\colhead{\textbf{Galaxy}} &
\colhead{\textbf{dist.}}&
\colhead{$\mathbf{\mathcal{M}_{gas}}$}&
\colhead{\textbf{M$_{V}$}} & 
\colhead{\textbf{M$_{H}$}} &
\colhead{\textbf{[Fe/H]}} & 
\colhead{\textbf{$h$}}}
\startdata
HIPASS\,J1112-86  & $29.09 \pm 4.36$ & $9.79\,(3.53)$\,E+08 &$-17.13 \pm 0.33$ &$-19.55 \pm 0.34$&$ -0.86 \pm 0.09$ & $1.13$\\  
HIPASS\,J0554-71  & $19.75 \pm 2.96$ & $4.64\,(1.67)$\,E+08  &$-15.06 \pm 0.34$ &$-17.50 \pm 0.35$&$ -1.21 \pm 0.07$ & $0.53$\\ 
HIPASS\,J1934-67  & $55.20 \pm 8.28$ & $3.62\,(1.31)$\,E+09 &$-19.29 \pm 0.33$ &$-21.31 \pm 0.33$&$ -0.48 \pm 0.12$ & $2.90$\\  
AM\,0433-654      & $16.60 \pm 2.49$ & $1.55\,(0.56)$\,E+08 &$-15.13 \pm 0.33$ &$-17.67 \pm 0.34$&$ -1.20 \pm 0.07$ & $1.29$\\  
IC\,5028          & $20.90 \pm 3.14$ & $1.40\,(0.50)$\,E+09 &$-16.90 \pm 0.34$ &$-18.59 \pm 0.34$&$ -0.90 \pm 0.08$ & $1.54$\\  
IC\,5008          & $50.02 \pm 7.50$ & $3.55\,(1.28)$\,E+09 &$-18.45 \pm 0.33$ &$-20.25 \pm 0.34$&$ -0.63 \pm 0.11$ & $1.98$\\  
ESO\,383-\,G\,092 & $19.15 \pm 2.87$ & $7.15\,(2.14)$\,E+08 &$-16.99 \pm 0.33$ &$-18.62 \pm 0.33$&$ -0.88 \pm 0.08$ & $0.43$\\ 
ESO\,318-\,G\,013 & $7.48  \pm 1.12$ & $1.75\,(0.63)$\,E+08&$-15.21 \pm 0.34$ &$-17.05 \pm 0.35$&$ -1.18 \pm 0.07$ & $0.90$\\  
HIPASS\,J1801-72  \tablenotemark{a}& $44.25 \pm 6.64$ & $3.88\,(1.40)$\,E+09&$-18.86 \pm 0.33$ &$-21.07 \pm 0.33$&$ -0.56 \pm 0.12$ & $0.77$\\  
ESO\,148-\,G\,006 & $42.95 \pm 6.44$ & $3.96\,(1.43)$\,E+09&$-18.15 \pm 0.34$ &$-20.18 \pm 0.34$&$ -0.68 \pm 0.10$ & $2.54$\\  
ESO\,084-\,G\,040 & $16.74 \pm 2.51$ & $6.11\,(2.20)$\,E+08&$-16.74 \pm 0.39$ &$-18.63 \pm 0.33$&$ -0.92 \pm 0.09$ & $0.90$\\  
HIPASS\,J1424-16b & $22.90 \pm 3.43$ & $2.27\,(0.82)$\,E+09&$-17.56 \pm 0.38$ &$-19.52 \pm 0.34$&$ -0.78 \pm 0.10$ & $1.64$\\  
HIPASS\,J0736-74  & $14.60 \pm 2.19$ & $1.55\,(0.59)$\,E+08&$-15.80 \pm 0.33$ &$-17.56 \pm 0.33$&$ -1.08 \pm 0.07$ & $0.44$\\  
ESO\,085-\,G\,088 & $15.66 \pm 2.35$ & $3.56\,(1.28)$\,E+08&$-17.08 \pm 0.38$ &$-19.28 \pm 0.42$&$ -0.86 \pm 0.09$ & $3.32$\\  
SGC\,0454.2-6138  & $12.93 \pm 1.94$ & $1.16\,(0.42)$\,E+08&$-14.75 \pm 0.35$ &$-16.56 \pm 0.34$&$ -1.26 \pm 0.07$ & $0.79$\\  
ESO\,052-\,G\,010 & $18.74 \pm 2.81$ & $4.41\,(1.59)$\,E+08&$-15.99 \pm 0.33$ &$-17.65 \pm 0.33$&$ -1.05 \pm 0.07$ & $0.72$\\  
ESO\,321-\,G\,014 & $3.19  \pm 0.48$ & $1.78\,(0.64)$\,E+07&$-12.55 \pm 0.33$ &$-14.87 \pm 0.35$&$ -1.64 \pm 0.10$ & $0.29$\\  
HIPASS\,J0653-73  & $15.53 \pm 2.33$ & $2.79\,(1.01)$\,E+08&$-14.56 \pm 0.34$ &$-17.09 \pm 0.34$&$ -1.30 \pm 0.07$ & $0.95$\\  
ESO\,140-\,G\,019 & $11.64 \pm 1.75$ & $1.88\,(0.68)$\,E+08&$-14.56 \pm 0.36$ &$-16.50 \pm 0.35$&$ -1.30 \pm 0.08$ & $0.93$\\  
HIPASS\,J0039-76  & $23.47 \pm 3.52$ & $6.91\,(2.49)$\,E+08&$-17.19 \pm 0.33$ &$-18.99 \pm 0.33$&$ -0.85 \pm 0.09$ & $0.66$\\  
\enddata
\tablecomments{Column (1) is the galaxy name; Column (2) is the
  adopted distance (in Mpc); Column (3) is the gas mass (in solar
  units); Column (4) is the absolute $V$-band magnitude; Column (5) is
  the absolute $H$-band magnitude; Column (6) is the calculated
  metallicity; Column (7) is the calculated mean scalelength (in kpc)
  of the $H$ and $V$-band exponential disks}
\tablenotetext{a}{Appears to be a double exponential disk, using inner disk for scalelength calculation}
\end{deluxetable}

\begin{deluxetable}{lcccccc}
\tablewidth{0pt}
\tabletypesize{\scriptsize}
\tablecaption{Evolutionary stellar population synthesis for the \textrm{H\,\scriptsize{I}} galaxies from ($V-H$) constrained simulations \label{tab:dis_mlt}}
\tablehead{
\colhead{\textbf{Galaxy}} &
\colhead{$(\mathcal{M}/L_V)$} &
\colhead{$(\mathcal{M}/L_H)$} & 
\colhead{$\mathbf{\mathcal{M}_{stars}}$:$V$}  &
\colhead{$\mathbf{\mathcal{M}_{stars}}$:$H$} & 
\colhead{$\mathbf{\mathcal{M}}$\textbf{\tiny{total}}:$V$} & 
\colhead{$\mathbf{\mathcal{M}}$\textbf{\tiny{total}}:$H$}} 
\startdata
HIPASS\,J1112-86   & $1.03 \pm 0.22$ &	$0.62 \pm 0.09$ & $6.75(2.53)$\,E\,+08 & $9.71(3.39)$\,E\,+08 &$1.65(0.55)$\,E\,+09 &$1.95(0.64)$\,E\,+09  \\
HIPASS\,J0554-71\tablenotemark{a}   & \nodata & \nodata & \nodata  & \nodata & \nodata & \nodata      						   \\
HIPASS\,J1934-67   & $0.76 \pm 0.13$ &	$0.46 \pm 0.06$ & $3.66(1.27)$\,E\,+09 & $3.60(1.19)$\,E\,+09 &$7.29(2.39)$\,E\,+09 &$7.23(2.34)$\,E\,+09  \\
AM\,0433-654       & $1.75 \pm 0.39$ &	$0.95 \pm 0.18$ & $1.82(0.69)$\,E\,+08 & $2.61(0.95)$\,E\,+08 &$3.37(1.14)$\,E\,+08 &$4.16(1.40)$\,E\,+08  \\
IC\,5028           & $0.82 \pm 0.19$ &	$0.55 \pm 0.09$ & $4.32(1.68)$\,E\,+08 & $3.52(1.23)$\,E\,+08 &$1.83(0.63)$\,E\,+09 &$1.75(0.60)$\,E\,+09  \\
IC\,5008           & $1.08 \pm 0.23$ &	$0.59 \pm 0.09$ & $2.38(0.88)$\,E\,+09 & $1.75(0.60)$\,E\,+09 &$5.93(1.99)$\,E\,+09 &$5.30(1.77)$\,E\,+09  \\
ESO\,383-\,G\,092\tablenotemark{b}  & \nodata & \nodata  & \nodata & \nodata & \nodata & \nodata        \\
ESO\,318-\,G\,013  & $0.58 \pm 0.04$ &	$0.49 \pm 0.02$ & $6.52(2.08)$\,E\,+07 & $7.65(2.47)$\,E\,+07 &$2.41(0.81)$\,E\,+08 &$2.52(0.84)$\,E\,+08  \\
HIPASS\,J1801-72   & $1.86 \pm 0.31$ &	$0.78 \pm 0.10$ & $6.01(2.07)$\,E\,+09 & $4.94(1.64)$\,E\,+09 &$9.90(3.23)$\,E\,+09 &$8.83(2.85)$\,E\,+09  \\
ESO\,148-\,G\,006  & $1.13 \pm 0.30$  &	$0.61 \pm 0.11$ & $1.89(0.77)$\,E\,+09 & $1.70(0.62)$\,E\,+09 &$5.85(2.00)$\,E\,+09 &$5.66(1.91)$\,E\,+09  \\
ESO\,084-\,G\,040  & $0.83 \pm 0.25$ &	$0.55 \pm 0.11$ & $3.76(1.75)$\,E\,+08 & $3.66(1.33)$\,E\,+08 &$9.87(3.47)$\,E\,+08 &$9.77(3.26)$\,E\,+08  \\
HIPASS\,J1424-16b  & $1.82 \pm 0.33$ &	$0.85 \pm 0.13$ & $1.77(0.69)$\,E\,+09 & $1.30(0.45)$\,E\,+09 &$4.04(1.37)$\,E\,+09 &$3.57(1.18)$\,E\,+09  \\
HIPASS\,J0736-74   & $0.61 \pm 0.11$ &	$0.49 \pm 0.06$ & $1.18(0.42)$\,E\,+08 & $1.22(0.40)$\,E\,+08 &$2.73(0.90)$\,E\,+08 &$2.77(0.90)$\,E\,+08  \\
ESO\,085-\,G\,088  & $1.05 \pm 0.34$ &	$0.62 \pm 0.13$ & $6.58(3.11)$\,E\,+08 & $7.59(3.32)$\,E\,+08 &$1.01(0.39)$\,E\,+09 &$1.12(0.42)$\,E\,+09  \\
SGC\,0454.2-6138   & $0.86 \pm 0.18$ &	$0.64 \pm 0.09$ & $6.24(2.39)$\,E\,+07 & $6.35(2.20)$\,E\,+07 &$1.78(0.60)$\,E\,+08 &$1.79(0.60)$\,E\,+08  \\
ESO\,052-\,G\,010  & $0.62 \pm 0.08$ &	$0.48 \pm 0.05$ & $1.42(0.47)$\,E\,+08 & $1.29(0.42)$\,E\,+08 &$5.83(1.97)$\,E\,+08 &$5.71(1.93)$\,E\,+08  \\
ESO\,321-\,G\,014  & $1.27 \pm 0.27$ &	$0.94 \pm 0.15$ & $1.23(0.46)$\,E\,+07 & $1.96(0.70)$\,E\,+07 &$3.02(1.01)$\,E\,+07 &$3.74(1.24)$\,E\,+07  \\
HIPASS\,J0653-73   & $1.51 \pm 0.08$ &	$0.89 \pm 0.04$ & $9.31(2.97)$\,E\,+07 & $1.43(0.45)$\,E\,+08 &$3.72(1.25)$\,E\,+08 &$4.22(1.39)$\,E\,+08  \\
ESO\,140-\,G\,019  & $1.73 \pm 0.27$ &	$1.01 \pm 0.14$ & $1.07(0.40)$\,E\,+08 & $9.51(3.35)$\,E\,+07 &$2.95(0.99)$\,E\,+08 &$2.83(0.95)$\,E\,+08  \\
HIPASS\,J0039-76   & $0.69 \pm 0.13$ &	$0.48 \pm 0.06$ & $4.79(1.70)$\,E\,+08 & $4.51(1.51)$\,E\,+08 &$1.17(0.39)$\,E\,+09 &$1.14(0.38)$\,E\,+09  \\ 
\enddata
\tablecomments{Column (1) is the galaxy name; Column (2) is the $V$-band $\mathcal{M}/L$  (in solar units); Column
  (3) is the $H$-band $\mathcal{M}/L$  (in solar units); Column (4) is the mass of stars (in solar units) using the stellar $V$-band  data; Column (5) is the mass of the stars (in solar units) using the $H$-band data; Column (6) is the $V$-band stellar plus gas mass (in solar units); Column (7) is the $H$-band stellar plus gas mass (in solar units)}
\tablenotetext{a}{Unable to model because no model ($V-H$) exists as red as the observed given the input metallicity and color}
\tablenotetext{b}{Ages younger than 8 Gyr}
\end{deluxetable}

\begin{deluxetable}{lcccccc}
\tabletypesize{\scriptsize}
\tablewidth{0pt}
\tablecaption{Derived baryon masses for the \textrm{H\,\scriptsize{I}} selected sample \label{tab:mas_den}}
\tablehead{
\colhead{\textbf{Galaxy}} &
\colhead{$\mathbf{\log W_{20}}$} &
\colhead{$\mathbf{\mathcal{F}}_{gas}$} &
\colhead{$\mathbf{\mathcal{M}}$\tiny{stars}} & 
\colhead{$\mathbf{\mathcal{M}}$\tiny{total}}  &
\colhead{$\mathbf{\log\Sigma_{stars}}$} & 
\colhead{$\mathbf{\log\Sigma_{total}}$}} 
\startdata
HIPASS\,J1112-86   &$2.11\,(0.04)$&$1.25\,(0.69)$&$7.81\,(2.03)$\,E+08&$1.78\,(0.42)$\,E+09&$1.99$&$2.35$\\
HIPASS\,J0554-71\tablenotemark{a}   &\nodata&\nodata&\nodata&\nodata&\nodata&\nodata \\                                                                                        
HIPASS\,J1934-67   &$2.37\,(0.05)$&$1.00\,(0.60)$&$3.63\,(0.87)$\,E+09&$7.26\,(1.67)$\,E+09&$1.84$&$2.14$\\
AM\,0433-654       &$1.82\,(0.12)$&$0.74\,(0.56)$&$2.09\,(0.56)$\,E+08&$3.68\,(0.88)$\,E+08&$1.30$&$1.55$\\
IC\,5028           &$2.19\,(0.05)$&$3.68\,(1.32)$&$3.80\,(0.99)$\,E+08&$1.79\,(0.43)$\,E+09&$1.41$&$2.08$\\
IC\,5008           &$2.28\,(0.05)$&$1.83\,(0.82)$&$1.95\,(0.50)$\,E+09&$5.58\,(1.32)$\,E+09&$1.90$&$2.35$\\
ESO\,383-\,G\,092\tablenotemark{b}  &\nodata&\nodata&\nodata&\nodata&\nodata&\nodata\\
ESO\,318-\,G\,013   &$1.89\,(0.04)$&$2.51\,(0.93)$&$6.99\,(1.59)$\,E+07&$2.46\,(0.58)$\,E+08&$1.13$&$1.68$\\
HIPASS\,J1801-72\tablenotemark{c}    &$2.47\,(0.02)$&$0.72\,(0.53)$&$5.36\,(1.28)$\,E+09&$9.29\,(2.14)$\,E+09&$3.16$&$3.40$\\                                                           
ESO\,148-\,G\,006   &$2.28\,(0.05)$&$2.23\,(0.97)$&$1.77\,(0.48)$\,E+09&$5.75\,(1.38)$\,E+09&$1.64$&$2.15$\\
ESO\,084-\,G\,040   &$2.08\,(0.07)$&$1.66\,(0.84)$&$3.69\,(1.06)$\,E+08&$9.82\,(2.38)$\,E+08&$1.86$&$2.28$\\
HIPASS\,J1424-16b   &$2.08\,(0.06)$&$1.58\,(0.77)$&$1.44\,(0.38)$\,E+09&$3.77\,(0.90)$\,E+09&$1.93$&$2.35$\\
HIPASS\,J0736-74    &$1.89\,(0.09)$&$1.29\,(0.67)$&$1.20\,(0.29)$\,E+08&$2.75\,(0.64)$\,E+08&$1.99$&$2.35$\\
ESO\,085-\,G\,088   &$1.91\,(0.05)$&$0.51\,(0.52)$&$7.05\,(2.27)$\,E+08&$1.06\,(0.29)$\,E+09&$1.01$&$1.18$\\
SGC\,0454.2-6138    &$1.87\,(0.09)$&$1.84\,(0.83)$&$6.30\,(1.62)$\,E+07&$1.79\,(0.42)$\,E+08&$1.21$&$1.66$\\
ESO\,052-\,G\,010   &$2.11\,(0.05)$&$3.27\,(1.12)$&$1.35\,(0.31)$\,E+08&$5.77\,(1.38)$\,E+08&$1.61$&$2.24$\\
ESO\,321-\,G\,014   &$1.65\,(0.14)$&$1.23\,(0.69)$&$1.45\,(0.39)$\,E+07&$3.31\,(0.78)$\,E+07&$1.44$&$1.80$\\
HIPASS\,J0653-73    &$2.05\,(0.05)$&$2.58\,(0.95)$&$1.08\,(0.25)$\,E+08&$3.95\,(0.93)$\,E+08&$1.28$&$1.84$\\
ESO\,140-\,G\,019   &$1.95\,(0.07)$&$1.88\,(0.84)$&$1.00\,(0.26)$\,E+08&$2.89\,(0.68)$\,E+08&$1.26$&$1.73$\\
HIPASS\,J0039-76    &$2.06\,(0.02)$&$1.49\,(0.72)$&$4.63\,(1.13)$\,E+08&$1.16\,(0.27)$\,E+09&$2.23$&$2.63$\\
\enddata            
\tablecomments{Column (1) is the galaxy name; Column (2) is the inclination corrected $W_{20}$ (in km\,s$^{-1}$); Column (3) is the
  gas--to--stars ratio ($\mathbf{\mathcal{M}_{gas}}$/$\mathbf{\mathcal{M}_{stars}}$); Column (4) is the stellar mass (in solar units); Column
  (5) is the gas plus stellar mass (in solar units); Column (6) is the mean
  stellar mass surface density (in $\mathcal{M}$$_{\odot}$\,Mpc$^{-2}$); Column (7)
  is the mean baryon mass surface density (in $\mathcal{M}$$_{\odot}$\,Mpc$^{-2}$)}
$^{a}$ and $^{b}$ same as in Tab. \ref{tab:dis_mlt}
\tablenotetext{c}{Appears to be a double exponential disk, using inner disk in mean baryon mass surface density calculation}
\end{deluxetable}
\clearpage
\begin{deluxetable}{lcccc}
\tablewidth{0pt}
\tabletypesize{\scriptsize}
\tablecaption{Parameters of the bivariate weighted fits \citep{pres92} for the TF/BTF relations with stellar $\mathcal{M}$/$L$ values calculated from population synthesis modelling and by using $\mathcal{M}$/$L$ values in braces, adopted from \cite{mcg2000}. \label{tab:tf_fits}}
\tablehead{
\colhead{\textbf{Sample}} &
\colhead{\textbf{TF slope}}&
\colhead{\textbf{BTF slope}} &
\colhead{\textbf{TF inter.}} & 
\colhead{\textbf{BTF inter.}}} 

\startdata
\textrm{H\,\scriptsize{I}} \& Sakai & $3.8 \,(3.8) \pm 0.1$ \,(0.1) & $3.2 \,(3.3) \pm 0.1$ \,(0.1) &$0.6 \,(1.0) \pm 0.3 \, (0.3)$ & $2.5 \, (2.4) \pm 0.3 \, (0.3)$ \\
\textrm{H\,\scriptsize{I}} & $3.4 \,(3.2) \pm 0.3$ \,(0.3) &  $3.0 \,(3.0) \pm 0.2$ \,(0.3)  &$1.5 \, (2.2) \pm 0.5 \,(0.6)$ & $2.8 \,(2.9) \pm 0.5 \,(0.6)$\\
Sakai & $4.3 \,(4.0) \pm 0.4$ \,(0.5) & $3.1 \,(3.2) \pm 0.3$ \,(0.3)   & $-0.6 \,(0.3)\, \pm 1.0 \,(1.2)$ & $2.6 \, (2.6) \pm 0.8 \, (0.9)$\\
\enddata
\tablecomments{Reduced Chi squared ($\mathbf{\chi^{2}_{red}}$) values for respective TF \& BTF bivariate weighted fits: $1.8 \, (1.4) \, \&\, 1.3 \,(1.2)\,, 2.5 \, (2.0) \, \&\, 1.7\, (1.5)\,,0.9\, (0.7)\, \&\, 0.9 \,(1.0)\,$}
\end{deluxetable}

\begin{deluxetable}{lccc}
\tablewidth{0pt} \tabletypesize{\scriptsize} \tablecaption{mean baryon mass surface density ($\Sigma_{b}$) values for different disk scalings. \label{tab:logsigma}}
\tablehead{ \colhead{\textbf{$\gamma$}} & \colhead{\textbf{1.0}}&
  \colhead{\textbf{1.5}} & \colhead{\textbf{2.0}}} \startdata
$\rm{R}_{\rm{e}}$/$h_{\star}$ & 1.68 & 2.04 & 2.33 \\ $\Sigma_{b}$ &
1.00 & 0.68 & 0.52 \\ log$\Sigma_{b}$ & 0.00 & -0.17 & -0.28
\\ \enddata \tablecomments{$\gamma$ values, defined as $h_{\textrm{H\,\scriptsize{I}}}$/$h_{\star}$, R$_{\rm{e}}$ is the half light radius,:
 Increasing the
  $h_{\textrm{H\,\scriptsize{I}}}$ scalelength for the dwarfs yields a
  $\log \Sigma_{b}$ \textit{vs} $\log W_{20}$ slope of $\sim 1.3$ which goes in
  the right sense to flatten the BTFR}
\end{deluxetable}

\begin{thebibliography}{}

\bibitem[{{Aaronson} \& {Mould} (1983)}]{aar83}
{Aaronson}, M., \& {Mould}, J. 1983, \apj, 265, 1

\bibitem[{{Allen} \& {Shanks}(2004)}]{allenandshanks2004}
{Allen}, P. D., \& {Shanks}, T. 2004, \mnras, 347, 1011

\bibitem[{{Barnes} {et~al.}(2001){Barnes}, {Staveley-Smith}, {de Blok},
  {Oosterloo}, {Stewart}, {Wright}, {Banks}, {Bhathal}, {Boyce}, {Calabretta},
  {Disney}, {Drinkwater}, {Ekers}, {Freeman}, {Gibson}, {Green}, {Haynes}, {te
  Lintel Hekkert}, {Henning}, {Jerjen}, {Juraszek}, {Kesteven}, {Kilborn},
  {Knezek}, {Koribalski}, {Kraan-Korteweg}, {Malin}, {Marquarding}, {Minchin},
  {Mould}, {Price}, {Putman}, {Ryder}, {Sadler}, {Schr{\"o}der}, {Stootman},
  {Webster}, {Wilson}, \& {Ye}}]{bar01}
{Barnes}, D.~G., {et~al.} 2001,
   \mnras, 322, 486

\bibitem[\protect\astroncite{{Battaglia} et~al.}{2005}]{bat05}
{Battaglia}, G., {Helmi}, A., {Morrison}, H., {Harding}, P., {Olszewski},
  E.~W., {Mateo}, M., {Freeman}, K.~C., {Norris}, J., \& {Shectman}, S.~A.
  2005, \mnras, 364, 433


\bibitem[{{Begum} \& {Chengalur}(2003)}]{beg2003}
{Begum}, A., \& {Chengalur}, J.~N. 2003, \aap, 409, 879

\bibitem[{{Begum} {et~al.}(2006) {Begum}, {Chengalur},{Karachentsev},{Kaisin}, \& {Sharina}}]{beg06}
{Begum}, A., {Chengalur}, J.~N., {Karachentsev}, I.~D., {Kaisin}, S.~S., \& {Sharina}, M.~E. 2006, \mnras, 365, 1220


\bibitem[{{Bell} \& {de Jong}(2000)}]{bel2000}
{Bell}, E.~F., \& {de Jong}, R.~S. 2000, \mnras, 312, 497

\bibitem[{{Bell} \& {de Jong}(2001)}]{bel01}
{Bell}, E.~F., \& {de Jong}, R.~S. 2001, \apj, 550, 212

\bibitem[{{Broeils} \& {van Woerden}(1994)}]{bro94}
{Broeils}, A.~H., \& {van Woerden}, H. 1994, \aaps, 107, 129

\bibitem[{{Bruzual} \& {Charlot}(2003)}]{bc03}
{Bruzual}, G., \& {Charlot}, S. 2003, \mnras, 344, 1000

\bibitem[{{Burkert}(2003)}]{burk2003}
{Burkert}, A. 2003, \apss, 284, 697

\bibitem[\protect\astroncite{{Bullock} et~al.}{2001}]{bul01}
{Bullock}, J.~S., {Dekel}, A., {Kolatt}, T.~S., {Kravtsov}, A.~V., {Klypin},
  A.~A., {Porciani}, C., \& {Primack}, J.~R. 2001, \apj, 555, 240

\bibitem[\protect\astroncite{{Cayatte} et~al.}{1990}]{cayateetal1990}
{Cayatte}, V., {van Gorkom}, J.~H., {Balkowski}, C., \& {Kotanyi}, C.: 1990, \aj, 100, 604

\bibitem[{{Courteau} {et~al.}(2003)}]{cou03}
{Courteau}, S., {MacArthur}, L.~A., {Dekel}, A., {van den Bosch}, F.,
{McIntosh}, D. \& {Dale}, D. 2003, astroph/0310440

\bibitem[\protect\astroncite{{Courteau} \& {Rix}}{1999}]{courteauandrix1999}
{Courteau}, S., \& {Rix}, H. 1999, \apj, 513, 561

\bibitem[{{Dalcanton} {et~al.}(2004){Dalcanton}, {Yoachim}, \&
  {Bernstein}}]{dalc04}
{Dalcanton}, J.~J., {Yoachim}, P., \& {Bernstein}, R.~A. 2004, \apj, 608, 189



\bibitem[\protect\astroncite{{de Vaucouleurs} et~al.}{1992}]{dev92}
{de Vaucouleurs}, G., {de Vaucouleurs}, A., {Corwin}, Jr., H.~G., {Buta},
  R.~J., {Paturel}, G., \& {Fouque}, P. 1992,{\em VizieR Online Data Catalog}, 7137, 0

\bibitem[{{Eder} \& {Schombert}(2000)}]{eder2k}
{Eder}, J.~A., \& {Schombert}, J.~M. 2000, \apjs, 131, 47

\bibitem[{{Ferrarese} {et~al.}(2000){Ferrarese}, {Ford}, {Huchra}, {Kennicutt},
  {Mould}, {Sakai}, {Freedman}, {Stetson}, {Madore}, {Gibson}, {Graham},
  {Hughes}, {Illingworth}, {Kelson}, {Macri}, {Sebo}, \& {Silbermann}}]{fer00}
{Ferrarese}, L., {et~al.} 2000, \apjs, 128, 431


\bibitem[{{Freeman}(1970)}] {kcf70}
{Freeman}, K.~C. 1970, \apj, 160, 811

\bibitem[{{Freeman}(1999)}]{kcf99}
{Freeman}, K.~C. 1999, in ASP Conf. Ser. 170:
  The Low Surface Brightness Universe, 3

\bibitem[{{Fukugita} \& {Peebles}(2006)}]{fuku06}
{Fukugita}, M., \& {Peebles}, P.J.E. 2006, \apj, 639, 590

\bibitem[{{Geha} {et~al.}(2006)}]{geha2006}
{Geha}, M., {Blanton}, M., {Masjedi}, M., \& {West}, A. 2006, \apj, 653, 240

\bibitem[{{Graham}(1982)}]{gra82}
{Graham}, J. A. PASP, 94, 244


\bibitem[{Gurovich {et~al.}(2004)Gurovich, {McGaugh}, {Freeman}, {Jerjen},
  {Staveley-Smith}, \& {De Blok}}]{guropasa}
Gurovich, S., {McGaugh}, S.~S., {Freeman}, K.~C., {et~al.} 2004, in  (Sydney: CSIRO), 412

\bibitem[\protect\astroncite{{Gurovich} et~al.}{2004}]{gurovichetal2004}
{Gurovich}, S., {McGaugh}, S.~S., {Freeman}, K.~C., {Jerjen}, H.,
  {Staveley-Smith}, L., \& {De Blok}, W.~J.~G. 2004 PASA, 21, 412


\bibitem[{Gurovich {}(2007)Gurovich}]{guroth}
Gurovich, S. 2007, Ph.D. thesis, Australian National University

\bibitem[{Gurovich {\& Freeman}(2007)Gurovich, {Freeman}}]{gurofree}
Gurovich, S., {Freeman}, K.~C. 2007, BAAA, 50, 251

\bibitem[{{Huchtmeier} \& {Richter}(1989)}]{huc89}
{Huchtmeier}, W.~K., \& {Richter}, O. -G. 1989, \aap, 210, 1

\bibitem[{{Jedrzejewski}(1987)}]{jed87}
{Jedrzejewski}, R.~I. 1987, \mnras, 226, 747

\bibitem[{{Kauffmann} {et~al.}(2003){Kauffmann}, {Heckman}, {White}, {Charlot},
  {Tremonti}, {Peng}, {Seibert}, {Brinkmann}, {Nichol}, {SubbaRao}, \&
  {York}}]{kau03}
{Kauffmann}, G., {et~al.} 2003, \mnras, 341, 54

\bibitem[{{Kassin} {et~al.}(2007){Kassin}, {Weiner}, {Faber}, {Koo}, {Lotz},
  {Diemand}, {Harker}, {Bundy}, {Metevier}, {Phillips}, {Cooper}, {Croton},
  {Konidaris}, {Noeske}, \& {Willmer}}]{kas2007}
{Kassin}, S.~A., {et~al.} 2007, \apjl, 660, L35

\bibitem[{{Kennicutt} {et~al.}(2003){Kennicutt}, {Armus}, {Bendo}, {Calzetti},
  {Dale}, {Draine}, {Engelbracht}, {Gordon}, {Grauer}, {Helou}, {Hollenbach},
  {Jarrett}, {Kewley}, {Leitherer}, {Li}, {Malhotra}, {Regan}, {Rieke},
  {Rieke}, {Roussel}, {Smith}, {Thornley}, \& {Walter}}]{ken03}
{Kennicutt}, R.~C., {et~al.} 2003, \pasp, 115, 928

\bibitem[{{Kilborn} {et~al.}(2002){Kilborn},{Webster},{Staveley-Smith}, 
	{Marquarding}, {Banks},  {Barnes},
	{Bhathal}, {de Blok},  {Boyce}, 
	{Disney}, {Drinkwater},{Ekers},
	{Freeman}, {Gibson},  {Henning},
	{Howlett}, {Jerjen}, {Knezek}, {Koribalski},
	{Malin}, {Minchin},  {Mould},{O'Brien},
	{Oosterloo}, {Ortiz}, {Price}, {Putman},  
	{Ryder},  {Sadler},  {Stewart}, 
	{Stootman}, {Wong}, \& {Wright}}]{kil2002} {Kilborn}, V.~A., {et~al.} 2002, \aj, 124, 690

\bibitem[{{Koribalski} {et~al.}(2004){Koribalski}, {Staveley-Smith}, {Kilborn},
  {Ryder}, {Kraan-Korteweg}, {Ryan-Weber}, {Ekers}, {Jerjen}, {Henning},
  {Putman}, {Zwaan}, {de Blok}, {Calabretta}, {Disney}, {Minchin}, {Bhathal},
  {Boyce}, {Drinkwater}, {Freeman}, {Gibson}, {Green}, {Haynes}, {Juraszek},
  {Kesteven}, {Knezek}, {Mader}, {Marquarding}, {Meyer}, {Mould}, {Oosterloo},
  {O'Brien}, {Price}, {Sadler}, {Schr{\"o}der}, {Stewart}, {Stootman}, {Waugh},
  {Warren}, {Webster}, \& {Wright}}]{kor2004}
{Koribalski}, B.~S., {et~al.} 2004, \aj, 128, 16

\bibitem[{{Kravtsov} {et~al.}(2004){Kravtsov},{Gnedin}, \& {Klypin}}]{krav04}
{Kravtsov}, A., {Gnedin}, O., \& {Klypin}, A. 2004, \apj, 609, 482

\bibitem[{{Lang} {et~al.}(2003){Lang}, {Boyce}, {Kilborn}, {Minchin}, {Disney},
  {Jordan}, {Grossi}, {Garcia}, {Freeman}, {Phillipps}, \& {Wright}}]{lan03}
{Lang}, R.~H., {et~al.} 2003, \mnras, 342, 738

\bibitem[{{Maller} \& {Bullock}(2004)}]{mall04}
{Maller}, A.~H., \& {Bullock}, J.~S. 2004, \mnras, 355, 694

\bibitem[{{Macri} {et~al.}(2000){Macri}, {Huchra}, {Sakai}, {Mould}, \&
  {Hughes}}]{mac2000}
{Macri}, L.~M., {Huchra}, J.~P., {Sakai}, S., {Mould}, J.~R., \& {Hughes},
  S.~M.~G. 2000, \apjs,
  128, 461

\bibitem[{{Martin}(1998)}]{mar98}
{Martin}, M.~C. 1998, \aaps, 131, 77

\bibitem[{{Mathewson} \& {Ford}(1996)}]{mat96}
{Mathewson}, D.~S., \& {Ford}, V.~L. 1996, \apjs, 107, 97

\bibitem[{{Meurer} {et~al.}(2009){Meurer}, {Wong}, {Kim}, {Hanish}, {Heckman},{Werk},{Bland-Hawthorn},{Dopita}, {Zwaan},{Koribalski}, {Seibert},{Thilker}, {Ferguson}, {Webster}, {Putman}, {Knezek}, {Doyle}, {Drinkwater}, {Hoopes}, {Kilborn}, {Meyer}, {Ryan-Weber}, {Smith}, \& {Staveley-Smith}}] {meureretal2009}
 {Meurer}, G. ~R., {et~al.} 2009, \apj, 695, 765

\bibitem[\protect\astroncite{{Meyer} et~al.}{2008}]{meyeretal2008}
{Meyer}, M.~J., {Zwaan}, M.~A., {Webster}, R.~L., {Schneider}, S., \&
  {Staveley-Smith}, L. 2008, \mnras, 391, 1712

\bibitem[{{McGaugh} {et~al.}(2000){McGaugh}, {Schombert}, {Bothun}, \& {de
  Blok}}]{mcg2000}
{McGaugh}, S.~S., {Schombert}, J.~M., {Bothun}, G.~D., \& {de Blok}, W.~J.~G.
  2000, \apjl, 533, L99

\bibitem[{{McGaugh}(2005)}]{McG2005}
{McGaugh}, S.~S. 2005, \apj,
  632, 859

\bibitem[{{Mateo}(1998)}]{mat98}
{Mateo}, M. 1998, ARA\&A, 36, 435

\bibitem[{{Mo} \& {Mao}(2000)}]{mo00}
{Mo}, H.~J., \& {Mao}, S. 2000, \mnras, 318, 163

\bibitem[{{Navarro} \& {Steinmetz}(2000)}]{nav00}
{Navarro}, J.~F., \& {Steinmetz}, M. 2000, \apj, 538, 477

\bibitem[{{Noordermeer} \& {Verheijen} (2007)}] {nor07}
{Noordermeer}, E., \& {Verheijen}, M. A. W. 2007, \mnras, 381, 1463

\bibitem[{{Pierce} \& {Tully}(1992)}]{pie92}
{Pierce}, M.~J., \& {Tully}, R.~B. 1992, \apj, 387, 47

\bibitem[{{Pildis} {et~al.}(1997){Pildis}, {Schombert}, \& {Eder}}]{pil97}
{Pildis}, R.~A., {Schombert}, J.~M., \& {Eder}, J.~A. 1997, \apj, 481, 157

\bibitem[{{Pilyugin} {et~al.}(2004){Pilyugin}, {V{\'{\i}}lchez}, \&
  {Contini}}]{pil04}
{Pilyugin}, L.~S., {V{\'{\i}}lchez}, J.~M., \& {Contini}, T. 2004, \aap, 425, 849

\bibitem[{{Press} {et~al.} (1992){Press},{Teukolsky},{Vettering} \& {Flannery}}]{pres92}
 {Press}, W. ~H., {Teukolsky}, S. ~A., {Vettering}, W. ~T., \& {Flannery}, B. ~P. 1992, Numerical Recipes in Fortran 77: The Art of Scientific Computing (2nd ed.; Cambridge: Cambridge Univ. Press), 660


\bibitem[{{Read} \& {Trentham}(2005)}]{rea2005}
{Read}, J.~I., \& {Trentham}, N. 2005, Royal Society of London Philosophical Transactions Series A,
  363, 2693

\bibitem[{{Rijcke} {et~al.}(2007)}]{rij2007}
{Rijcke}, S., {Zeilinger}, W., {Hau}, G., {Prugniel}, P., \&
{Dejonghe}, H. 2007, \apj, 659, 1172

\bibitem[{{Sakai} {et~al.}(2000){Sakai}, {Mould}, {Hughes}, {Huchra}, {Macri},
  {Kennicutt}, {Gibson}, {Ferrarese}, {Freedman}, {Han}, {Ford}, {Graham},
  {Illingworth}, {Kelson}, {Madore}, {Sebo}, {Silbermann}, \&
  {Stetson}}]{sak2000}
{Sakai}, S., {et~al.} 2000, \apj, 529, 698

\bibitem[{{Sandage}(1973)}]{san73}
{Sandage}, A. 1973, \apj, 183, 711

\bibitem[{{Schlegel} {et~al.}(1998){Schlegel}, {Finkbeiner}, \&
  {Davis}}]{sch98}
{Schlegel}, D.~J., {Finkbeiner}, D.~P., \& {Davis}, M. 1998, \apj, 500, 525


\bibitem[{{Schr{\"o}der} {et~al.}(2001){Schr{\"o}der}, {Drinkwater}, \&
  {Richter}}]{sch01}
{Schr{\"o}der}, A., {Drinkwater}, M.~J., \& {Richter}, O.-G. 2001,
  \aap, 376, 98

\bibitem[{{Schombert}(2006)}]{sch06}
{Schombert}, J.~M. 2006, \aj, 131, 296

\bibitem[\protect\astroncite{{Stark} et~al.}{2009}]{starketal2009}
{Stark}, D.~V., {McGaugh}, S.~S., \& {Swaters}, R.~A. 2009,
\newblock \aj, 138, 392

\bibitem[{{Swaters} {et~al.}(2003){Swaters},{Verheijen},{Bershady}, \& {Andersen}}]{swa2003}
{Swaters}, R.~A., {Verheijen}, M.~A.~W., {Bershady}, M.~A., \& 
	{Andersen}, D.~R. 2003, \apjl, 587, L19

\bibitem[{{Tassis} {et~al.}(2008){Tassis}, {Kravtsov}, \& {Gnedin}}]{tas2008}
{Tassis}, K., {Kravtsov}, A.~V., \& {Gnedin}, N.~Y. 2008, \apj, 672, 888

\bibitem[{{Tifft} \& {Huchtmeier}(1990)}]{tif90}
{Tifft}, W.~G., \& {Huchtmeier}, W.~K. 1990, \aaps, 84, 47

\bibitem[{{Trachternach} {et~al.}(2009){de Blok}, {McGaugh}, {van der Hulst},
  \& {Dettmar}}]{tra09}
{Trachternach}, C., {de Blok}, W.~J.,~G., {McGaugh}, S.~S., {van der Hulst}, J.~M., and {Dettmar}, R .-J. 2009, \aap, 505, 507

\bibitem[{{Tremonti} {et~al.}(2004){Tremonti}, {Heckman}, {Kauffmann},
  {Brinchmann}, {Charlot}, {White}, {Seibert}, {Peng}, {Schlegel}, {Uomoto},
  {Fukugita}, \& {Brinkmann}}]{tre04}
{Tremonti}, C.~A., {et~al.} 2004, \apj, 613, 898

\bibitem[{{Vallejo} {et~al.}(2002){Vallejo}, {Braine}, \& {Baudry}}]{val02}
{Vallejo}, O., {Braine}, J., \& {Baudry}, A. 2002, \aap, 387, 429

\bibitem[{{van den Bosch}(2000)}]{vand00}
{van den Bosch}, F.~C. 2000, \apj, 530, 177

\bibitem[{{Verheijen}(2001)}]{ver2001}
{Verheijen}, M.~A.~W. 2001, \apj, 563, 694

\bibitem[{{White}(1997)}]{whi97}
{White}, S. D. M. 1997, Galaxy Scaling Relations: Origins, Evolution and
  Applications, ed. L.~N. da Costa \& A.~Renzini (New York: Springer), 3

\bibitem[{{Zwaan} {et~al.}(1995){Zwaan}, {van der Hulst}, {de Blok}, \&
  {McGaugh}}]{zwa95}
{Zwaan}, M.~A., {van der Hulst}, J.~M., {de Blok}, W.~J.~G., \& {McGaugh},
  S.~S. 1995, \mnras, 273, L35
\end{thebibliography}
\end{document}